\documentclass[11pt,a4paper]{article}
\pdfoutput=1
\usepackage[pdftex]{graphics}
\usepackage{jheppub}
\usepackage{amsmath,amssymb,amsfonts}
\usepackage{array,booktabs}
\usepackage{slashed}
\usepackage[force]{feynmp-auto}
\usepackage{cancel}
\usepackage{xcolor}
\usepackage{kotex}

\usepackage{tikz}
\usepackage{tikz-feynman} 
\usetikzlibrary{decorations.markings, shapes.symbols}
\usepackage{bbding} 
\definecolor{cutbrown}{RGB}{176,84,45}   
\definecolor{momblue}{RGB}{40,80,180}    

\newcommand{\be}{\begin{eqnarray}}
\newcommand{\ee}{\end{eqnarray}}

\newcommand{\bn}{\begin{enumerate}}
\newcommand{\en}{\end{enumerate}}
\newcommand{\bl}{\begin{align}}
\newcommand{\el}{\end{align}}

\def\CC{{\cal C}}

\def\CI{{\cal I}}

\def\CK{{\cal K}}

\def\CO{{\cal O}}

\def\CZ{{\cal Z}}

\def\a{\alpha}
\def\b{\beta}
\def\g{\gamma}

\def\e{\epsilon}

\def\z{\zeta}
\def\th{\theta}

\def\k{\kappa}
\def\l{\lambda}
\def\m{\mu}
\def\n{\nu}
\def\r{\rho}

\def\s{\sigma}

\def\t{\tau}

\def\w{\omega}

\def\D{\Delta}

\def\L{\Lambda}

\def\para{\parallel}

\def\p{\partial}

\def\Tr{{\rm Tr}}

\def\da{{\dot{\a}}}

\def\jmath{{j}}

\def\d{{\delta}}
\def\be{{\bar{\epsilon}}}
\def\bs{{\bar{\sigma}}}

\def\hem{\hspace{0.05em}}

\usepackage{graphicx}
\usepackage[export]{adjustbox}

\usepackage{bm}

\def\rmx{{
    \scalebox{1.2}[1]{$\mathrm{x}$}\kern-0.625em\scalebox{1.2}[1]{$\mathrm{x}$}
}}
\def\rmy{{
    \scalebox{1.2}[1]{$\mathrm{y}$}\kern-0.645em\scalebox{1.2}[1]{$\mathrm{y}$}\kern-0.02em
}}
\def\rmz{{
    \kern0.055em\scalebox{1.2}[1]{$\mathrm{z}$}\kern-0.528em\scalebox{1.2}[1]{$\mathrm{z}$}\kern0.04em
}}

\newcommand{\dbar}{
    d\kern-.20em\makebox[0pt][l]{$\bar{}$}\kern.20em
}
\newcommand{\deltabar}{
    \delta\kern-.20em\makebox[0pt][l]{$\bar{}$}\kern.20em
}

\usepackage{mathtools}

\newcommand{\Ket}[1]{{\hem\big|\hem{#1}\big\rangle}}

\newcommand{\lambdabar}{
    \lambda\kern-.20em\makebox[0pt][l]{$\bar{}$}\kern.20em
}

\def\M{{\smash{M^\star}\vphantom{M}}}

\def\rg{\rangle}
\def\lg{\langle}

\def\fq{\texttt{q}}
\def\bfq{\bar{\texttt{q}}}
\def\fb{\texttt{b}}
\def\bfb{\bar{\texttt{b}}}
\def\fd{\texttt{d}}
\def\bfd{\bar{\texttt{d}}}
\def\fa{\texttt{a}_\bot}
\def\bfa{\bar{\texttt{a}}_\bot}

\title{Gravitational Faraday rotation, gravitational spin Hall effect, and spin-refined causality analysis from Magnusian matrix in effective field theories of gravity} 

\author[a]{Kibok Jeong}
\author[b]{Jung-Wook Kim}
\author[a]{Soochang Lee} 

\affiliation[a]{Department of Physics and Astronomy $\&$ Center for Theoretical Physics,
		\\
		Seoul National University, 1 Gwanak-ro, Seoul 08826, Korea}

\affiliation[b]{Theoretical Physics Department, CERN,
	1211 Geneva 23, Switzerland}

\abstract{
The effects of black hole's spin in effective field theories (EFTs) of gravity are explored through observables of wave scattering on black hole backgrounds in the geometric optics approximation. 
The considered observables are polarisation rotation angle, wavenumber kick (or deflection angle), and (Shapiro/Wigner--Smith) time delay, each of which are related to gravitational Faraday rotation, gravitational spin Hall effect, and infrared causality.
The observables are computed from scattering amplitudes through the Magnusian formalism, where the Magnusian is promoted to a matrix to account for helicity information. 
It is found that (1) the gravitational spin Hall effect in EFTs of gravity are qualitatively different from that of general relativity due to ``noncommutative'' wavenumber kicks, and (2) black hole's spin slightly enhances the causality constraints on EFT coefficients. 
}

\emailAdd{boki0322@snu.ac.kr}
\emailAdd{jung-wook.kim@cern.ch}
\emailAdd{physicsmp1217@snu.ac.kr}

\begin{document}
\begin{flushright}
\vspace{10pt} \hfill{CERN-TH-2026-190} \vspace{20mm}
\end{flushright}
\maketitle

\section{Introduction}
The first approximation employed when studying black holes in effective field theories of gravity is spherical symmetry, which leads to Schwarzschild-like non-spinning black hole solutions. 
However, practically all black holes observed in nature are spinning; we do not know of a realistic process that would result in a black hole with exactly zero angular momentum. 
What are the physical effects induced by black hole spins that are absent in non-spinning black holes in effective field theories (EFTs) of gravity? 
Can we use black hole spins as a tool to improve our understanding of the UV completion of general relativity?

We answer this question using scattering amplitudes, specialising to the case of wave scattering on black hole spacetime. 
This approach has the advantage that black hole solutions in modified gravity theories are not necessary, which can become almost impossible to obtain analytically when spherical symmetry is lost in modified Einstein's equations. 
The scattering of waves on black hole backgrounds can be studied using $2 \to 2$ massive--massless scattering amplitudes, where the massive particle plays the role of the black hole and the massless particle corresponds to the wave that is scattered by the black hole background. 
The scattering observables are computed using the Magnusian formalism~\cite{Kim:2025gis}, which can be viewed as a refinement of the eikonal approximation of $2 \to 2$ scattering amplitudes. 
We generalise the Magnusian to Magnusian matrix since polarisation plays an important role in our analysis, similar to how the eikonal phase had to be generalised to eikonal phase matrix in ref.~\cite{AccettulliHuber:2020oou}. 

Black hole's spin induces the following effects that are absent in non-spinning black holes: gravitational Faraday rotation, gravitational spin Hall effect, and spin refinement of (Shapiro) time delay. 
We study these effects when the Einstein--Maxwell action is modified by $FFR$, $R^3$, and $R^4$ corrections, where the modifications are treated as EFT corrections. 
How to encode \emph{classical} spins of black holes as \emph{quantum} spins of ``elementary'' particles in scattering amplitudes has been well documented in the post-Minkowskian dynamics literature~\cite{Guevara:2018wpp,Chung:2018kqs,Damgaard:2019lfh,Bern:2020buy,Maybee:2019jus,Aoude:2020onz,Bautista:2021wfy,Aoude:2021oqj,Cangemi:2022bew,Bjerrum-Bohr:2023jau,Kim:2023drc,Haddad:2023ylx,Bern:2023ity,Luna:2023uwd,Akpinar:2024meg,Vazquez-Holm:2025ztz}; we use the treatment of ref.~\cite{Chen:2021kxt} in this work, which is also the approach used to study gravitational Faraday rotation in general relativity using scattering amplitudes~\cite{Chen:2022clh,Kim:2022iub}.

This paper is organised as follows. 
Sec.~\ref{sec:Magnusian} gives a brief overview of the considered physical effects (gravitational Faraday rotation, gravitational spin Hall effect, and time delay) and how to compute them in the Magnusian formalism. 
Sec.~\ref{sec:setup} provides details of the setup such as the EFT action and scattering kinematics.
Sec.~\ref{sec:observables1} summarises the gravitational Faraday rotation and gravitational spin Hall effect computed using the Magnusian formalism, and sec.~\ref{sec:TD_causality} summarises the time delay and causality analysis based on the time delay.
We conclude the paper with discussions in sec.~\ref{sec:conclusions}.

\section{Probing physics using observables from the Magnusian matrix} \label{sec:Magnusian}
\subsection{From the Magnusian to the Magnusian matrix}

The \emph{Magnusian} $\chi$ is defined as the log of the $S$-matrix $\chi = \frac{\hbar}{i} \log S$; its matrix elements are also known as \emph{Magnus amplitudes}~\cite{Kim:2025gis,Brandhuber:2025igz}. 
The Magnusian was introduced as an attempt to formalise the eikonal approximation of $2 \to 2$ scattering amplitudes through the exponential representation of the $S$-matrix $S = e^{i \chi/\hbar}$~\cite{Damgaard:2021ipf,Damgaard:2023ttc}. 
Although it was eventually understood that the Magnusian $\chi$ is \emph{not equivalent} to the eikonal phase $\delta$ of the eikonal approximation~\cite{Kim:2025gis,Kim:2025sey},\footnote{\label{fn:eik_Mag}The eikonal approximation should be understood as the saddle-point approximation of the $S$-matrix element; see ref.~\cite{DiVecchia:2023frv} for a review. The distinction between the \emph{physical impact parameter} $b_J$ and the \emph{eikonal impact parameter} $b_\text{eik}$ arises from perturbative corrections to the saddle point, which must be considered at one-loop order for spinning systems when computing scattering observables~\cite{Luna:2023uwd}. This distinction is unnecessary in the Magnusian approach, where the nested brackets of the scattering generator equation \eqref{eq:scgendef} emulates the effects of the saddle-point shift~\cite{Kim:2025hpn}.} they are numerically the same at sufficiently low orders in perturbation theory, and we will compute the Magnusian as the eikonal phase.\footnote{In principle, it is possible to compute the Magnusian directly from modified diagrammatic rules~\cite{Brandhuber:2025igz,Kim:2025ebl,Guo:2026xaw}; see e.g. ref.~\cite{Bautista:2026fcp}. We do not pursue this computation method for practical reasons: (1) it is not fully established whether modern unitarity methods~\cite{Bern:1994zx,Bern:1994cg,Bern:1997sc} can be applied directly to loop-level Magnus amplitudes, and (2) the relevant master integrals with modified $i0^+$ prescriptions are not easily accessible in the literature.} 

In the Magnusian approach, the phase-space formulation of quantum mechanics is adopted~\cite{Kim:2025ebl} and scattering observables are computed using (nested) Poisson brackets\footnote{The Poisson brackets suffice since we only focus on the (appropriately defined) leading $\hbar$ contributions, where the coefficients of the EFT operators are normalised by appropriate powers of $\hbar$ such that the leading effects scale as classical contributions; see sec.~\ref{sec:geoopt}. The brackets need to be deformed to Moyal brackets when subleading $\hbar$ contributions become relevant.} through the \emph{scattering generator equation}~\cite{Gonzo:2024zxo,Kim:2024grz,Kim:2024svw,Kim:2025hpn,Kim:2025gis,Brandhuber:2025igz},
\begin{align}
    O_{\text{out}} &= e^{\{ \chi, \bullet \}} [O_{\text{in}}] = O_{\text{in}} + \{ \chi, O_{\text{in}} \} + \frac{1}{2!} \{ \chi , \{ \chi , O_{\text{in}} \} \} + \cdots \,, \label{eq:scgendef}
\end{align}
which can be viewed as a symmetry transform generated by $\chi$. The consideration of nested Poisson brackets is necessary when scattering observables are computed beyond the leading order. 
Since we are working up to one-loop order, we will only need to consider up to the double bracket term of \eqref{eq:scgendef}.

The fundamental Poisson brackets are ($P_1^\m = p_1^\m$ massive momentum, $P_2^\m = k_2^\m$ massless momentum, mostly-negative metric, $\e^{0123} = +1$, covariant spin supplementary condition)~\cite{Steinhoff:2015ksa,Kim:2026bqi}
\begin{align}
\begin{aligned}
    \{ X_i^\m , P_j^\n \} = - \delta_{ij} \eta^{\m\n} \,&,\quad \{ X_1^\m , S^\n \} = \frac{p_1^\m S^\n + S^\m p_1^\n - \eta^{\m\n} (p_1 \cdot S)}{m^2} \,,
    \\ \{ S^\m , S^\n \} = - \frac{1}{m} \epsilon^{\m\n\a\b} p_{1\a} S_{\b} \,&,\quad \{ X_1^\m , X_1^\n \} = - \frac{1}{m^3} \epsilon^{\m\n\a\b} p_{1\a} S_{\b} \,.
\end{aligned}
\end{align}
Generally nested Poisson brackets do \emph{not} vanish because the definition of the impact parameter should manifestly satisfy the orthogonality conditions $(b \cdot p_1) = (b \cdot k_2) = 0$~\cite{Kim:2024grz,Kim:2025hpn}; the impact parameter should be defined as
\begin{align}
    b^\mu &:= X_{21}^\m - (X_{21} \cdot p_1) \frac{k_2^\m}{p_1 \cdot k_2} - (X_{21} \cdot k_2) \left[ \frac{p_1^\m}{p_1 \cdot k_2} - \frac{m^2 k_2^\m}{(p_1 \cdot k_2)^2} \right] \,,
\end{align}
where $X_{21}^\m := X_2^\m - X_1^\m$, $X^\m_i$ is the position variable associated with the particle $i$ in the classical interaction picture~\cite{Kim:2025olv}, the index 1 is associated with the massive particle $p_1^2 = m^2$ identified as the Kerr background, and the index 2 is associated with the massless particle (photon or graviton). The impact parameter Poisson brackets can be written as
\begin{align}
\begin{aligned}
    \{ b^\m , b^\n \} &\stackrel{\cdot}{=} - \frac{(m + \w)\left( b^\m k_2^\n - k_2^\m b^\n \right)}{m \w^2} + \frac{ b^\m p_1^\n - p_1^\m b^\n }{m \w}
    \\ & \phantom{=as} - \frac{\e^{\m\n\a\b}p_{1\a}S_\b}{m^3} + \frac{2p_1^{[\m}\e^{\n]\a\b\g}p_{1\a}k_{2\b}S_\g}{m^4 \w} - \frac{2k_2^{[\m}\e^{\n]\a\b\g}p_{1\a}k_{2\b}S_\g}{m^3 \w^2} \,,
\end{aligned}
\end{align}
where $\stackrel{\cdot}{=}$ indicates equivalence up to on-shell conditions and $\w = \frac{p_1 \cdot k_2}{m}$. The terms on the second line are irrelevant for the $\hbar$ expansions considered in this work.

\paragraph{Magnusian matrix}
When we take the polarisation information of the massless particle into account, the eikonal phase $\delta$ is promoted to the eikonal phase matrix $\overleftrightarrow{\delta}$ where $\delta_{IJ}$ are its matrix elements and $I,J = \pm$ is the helicity information~\cite{AccettulliHuber:2020oou}. Analogously, the Magnusian $\chi$ should be promoted to the Magnusian matrix $\overleftrightarrow{\chi}$. To define its action $\{ \overleftrightarrow{\chi} , \bullet \}$, we promote an observable $x$ to a vector-valued observable $\vec{x}$ in helicity space, where $x_I$ are its components. We assign $x_+ = x_- = x$ if the observable $x$ does not depend on the polarisation of the massless particle; we assign $x_+ = x$ and $x_- = 0$ ($x_- = x$ and $x_+ = 0$) if the observable is only defined for massless particle's positive (negative) helicity state.\footnote{Effectively, we are considering observables which can be simultaneously diagonalised with helicity and can be used as ``good quantum numbers'' for specifying the state in the free theory.} We define the action of the Magnusian matrix $\overleftrightarrow{\chi}$ as
\begin{align}
    \vec{O}' = \{ \overleftrightarrow{\chi} , \vec{O} \} \quad \Rightarrow \quad O'_I = \sum_{J = \pm} \{ \chi_{IJ} , O_J \} \,,
\end{align}
which makes the linear operator property of $\{ \overleftrightarrow{\chi} , \bullet \}$ manifest. To avoid clutter, we will simply write the Magnusian matrix as $\chi$ or as $\chi_{IJ}$, the latter when its matrix nature needs to be explicitly stated.

The expectation value of an observable is computed (in the classical limit) as 
\begin{align}
    \Tr \left[ \rho \; e^{\frac{1}{i\hbar} [\chi , \bullet]}[O] \right] \; \stackrel{\hbar \to 0}{\to} \; \Tr \left[ \rho \; e^{\{ \chi , \bullet \}}[O] \right] \,,
\end{align}
where $\rho$ is the density matrix corresponding to the initial helicity configuration of the massless particle. When the matrix $e^{\{ \chi , \bullet \}}[O]$ is diagonalised, one can prepare an initial pure/coherent state where one of its diagonal elements becomes the experimental outcome.

\subsection{Polarisation rotation angle and gravitational Faraday rotation}
\begin{figure}[h]
    \centering
    \includegraphics[width=0.4\textwidth]{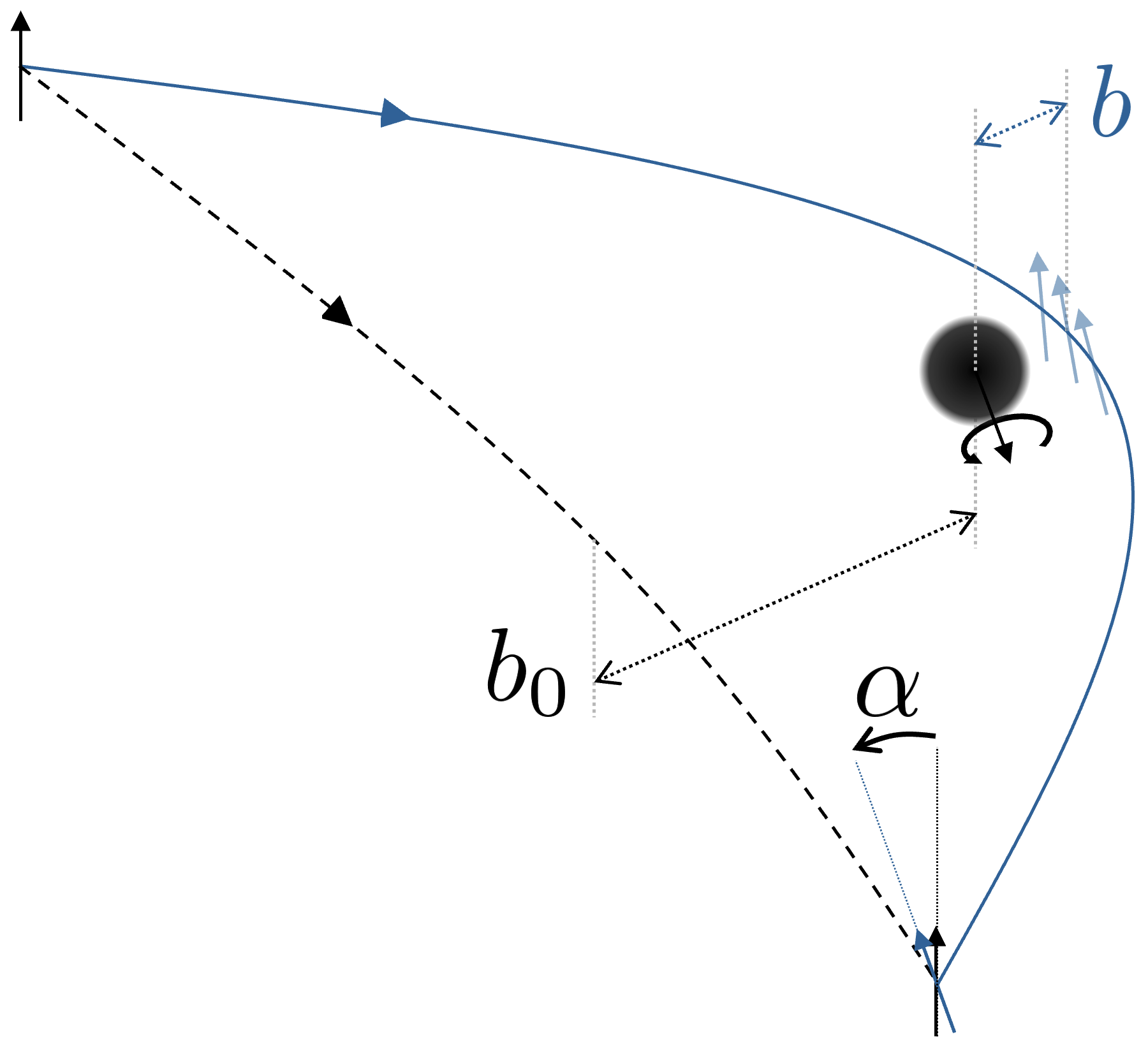}
    \caption{Experimental setup for measuring gravitational Faraday rotation. The trajectory that passes by the black hole has impact parameter $b$, while the trajectory that is unaffected by the black hole's gravitational field has impact parameter $b_0$. The limit $b_0 \to \infty$ is computed by \eqref{Pol rotation angle}. A similar setup can be used to measure the wavenumber kick \eqref{def: Deflection angle} and the time delay \eqref{def: Time delay}.
    }
    \label{fig : Def of observable}
\end{figure}
Faraday rotation (Faraday effect) is a magneto-optic effect where the linear polarisation of light in a dielectric medium rotates due to the magnetic field applied parallel to its propagating direction~\cite{landau1984electrodynamics}.
The same phenomenon occurs in gravitomagnetic backgrounds, which is known as the gravitational Faraday rotation~\cite{Dehnen:1973xa}. 
This phenomenon can be understood as a circular birefringence of the vacuum induced by the frame-dragging field. 
In the geometric optics approximation, the effect can be computed using parallel transport along null geodesics~\cite{Ishihara:1987dv} or using the eikonal approximation of scattering amplitudes~\cite{Chen:2022clh}. 

The definition of the polarisation rotation angle based on the eikonal phase matrix~\cite{Chen:2022clh} extends naturally to the Magnusian matrix,
\begin{align} \label{Pol rotation angle}
    \a &:= \frac{\chi_{-\to-} - \chi_{+\to+}}{2h} \,,
\end{align}
where $h$ is the helicity of the massless particle. Experimentally, the rotation angle is measured by comparing the polarisation directions of light/gravitational waves emanating from a common source that followed different trajectories, one passing nearby a black hole under the influence of its gravity (impact parameter $b$) and the other sufficiently far away that can be considered as freely propagating (impact parameter $b_0$)~\cite{Ishihara:1987dv}; see fig.~\ref{fig : Def of observable}. 
The adopted definition for the polarisation rotation angle \eqref{Pol rotation angle} computes the limit where $b_0 \to \infty$.

\subsection{Wavenumber kick and gravitational spin Hall effect}
The trajectories of probe particles on black hole backgrounds are given by geodesics at the leading order approximation, but the trajectories can be corrected by the particle's spin degrees of freedom. 
This phenomenon is known as the gravitational spin Hall effect; see ref.~\cite{Oancea:2019pgm} for a review. 
We study the gravitational spin Hall effect of light and gravitational waves in effective field theories of gravity through scattering amplitudes. 

The gravitational spin Hall effect can be computed as polarisation dependence of the wavenumber kick (the massless equivalent of the impulse) using scattering amplitudes, where the impulse becomes a matrix-valued observable due to polarisation dependence. 
Since the black hole spin breaks the reflection symmetry in the impact parameter space, there are non-vanishing impulse components orthogonal to the impact parameter and the independent components of the impulse must be considered at the same time. 
We find that the two independent components are not simultaneously diagonalisable and leads to the ``noncommutativity'' of the impulse, similar to how the independent components of the spin variables cannot be simultaneously diagonalised. 
This implies that unlike the Stern--Gerlach experiment, the incoming ray will not split into two distinguishable rays---each ray associated to mutually orthogonal spin states of photons or gravitons---but rather disperse and become a blob of rays. 
To the best of our knowledge, such an effect has not been reported in the literature before. 

We compute the wavenumber kick $\D k_2^\m = k_3^\m - k_2^\m$ from the scattering generator equation \eqref{eq:scgendef}. 
We work up to one-loop order and truncate the exponentiated brackets at the quadratic order
\begin{align} \label{def: Deflection angle}
    k_3^\m = e^{\{ \chi , \bullet \}} [k_2^\m] = k_2^\m + \{ \chi , k_2^\m \} + \frac{1}{2!} \{ \chi , \{ \chi , k_2^\m \} \} \,.
\end{align}
The initial wavenumber $k_2^\m$ is associated with the direction of the incoming ray, while the final wavenumber $k_3^\m$ is associated with the direction of the outgoing ray.

\subsection{Generalised time delay and causality analysis}

One of the main tenets of physics is causality---the effect cannot precede the cause. 
While analyticity of the $S$-matrix is commonly regarded as an equivalent condition of causality~\cite{Eden:1966dnq}, analyticity per se is not an observable that can be measured in experiments.

An observable widely regarded as a probe of causality is the (Shapiro/Wigner--Smith) time delay~\cite{Wigner:1955zz,Smith:1960zza,Shapiro:1964uw}. 
Several arguments relate negative time delay (or time \emph{advance}) to violations of conditions implied by causality, such as the violation of the Gao--Wald theorem or the possibility of forming closed time-like curves~\cite{Adams:2006sv,Camanho:2014apa,Lee:2026rtz}.\footnote{It is also possible to understand the positivity of the time delay as a consequence of analyticity~\cite{Bellazzini:2022wzv}.}

We \emph{define} the time delay of the massless particle, $\D t$, as
\begin{align} \label{def: Time delay}
    \D t := \left( e^{\{ \chi , \bullet \}} - 1 \right) [X_2^0] = \{ \chi , X_2^0 \} + \frac{1}{2!} \{ \chi , \{ \chi , X_2^0 \} \} + \cdots \,.
\end{align}
This definition can be justified by considering a worldline description of particles in the classical interaction picture~\cite{Kim:2025olv}. 
When a particle is described using worldlines such as in worldline quantum field theory~\cite{Mogull:2020sak}, $X^0$ corresponds to the ``time'' coordinate of the particle at $\s = 0$ when interactions are turned off, where $\s$ is the affine parameter for the (free-propagating) worldline. 
The scattering generator equation \eqref{eq:scgendef} applied to $X^0$, \eqref{def: Time delay}, computes the change in $X^0$ before and after scattering, which can be interpreted as the translation of the free-propagating worldline in the temporal direction due to interactions.

Writing the time component of $k_2^\m$ as $k_2^0 = \w$, the definition \eqref{def: Time delay} is equivalent to
\begin{align} \label{eq: time delay as energy derivative}
    \D t &= \frac{\partial \chi}{\partial \w} + \frac{1}{2!} \{ \chi , \frac{\partial \chi}{\partial \w} \} + \cdots \,.
\end{align}
If we neglect the subleading Poisson brackets and identify the Magnusian as the eikonal phase $\chi = \delta$, this definition of the time delay reduces to the usual definition ($\D t = \frac{\partial \delta}{\partial \w}$) found in the literature~\cite{Camanho:2014apa,AccettulliHuber:2020oou}, which is only well-defined for $2 \to 2$ scattering since the eikonal approximation is only known for $2 \to 2$ kinematics. The Magnusian is well-defined even for multiparticle kinematics, e.g. $3 \to 3$ scattering, thus this definition of the time delay can be understood as a multiparticle generalisation of the original definition. We also remark that \eqref{eq: time delay as energy derivative} can also be understood as the (modified)\footnote{The lifetime matrix $Q$ defined by Smith is $Q = - i \hbar S \frac{\partial S^\dagger}{\partial E}$~\cite{Smith:1960zza}.} Smith time delay $\D t = - i \hbar S^\dagger \frac{\partial S}{\partial E}$ in the classical limit: from the derivative of the exponential map,
\begin{align}
    \frac{d}{dt} e^X &= e^X \; \frac{1 - e^{- \text{ad}_X}}{\text{ad}_X} \; \frac{dX}{dt} \,,
\end{align}
where $\text{ad}_X Y = [X,Y]$ is the adjoint action, we can evaluate the Smith time delay operator using the exponential form of the $S$-matrix, $S = e^{i \chi / \hbar}$, as
\begin{align}
\begin{aligned}
    - i \hbar e^{- i \chi / \hbar} \frac{\partial}{\partial E} e^{i \chi / \hbar} = \frac{e^{\frac{1}{i \hbar} [ \chi , \bullet]} - 1}{\frac{1}{i \hbar} [ \chi , \bullet]} \frac{\partial \chi}{\partial E} &= \sum_{k=0}^\infty \frac{1}{(k+1)!} \left( \frac{1}{i \hbar} [\chi , \bullet] \right)^{k} \left[ \frac{\partial \chi}{\partial E} \right]
    \\ & \; \stackrel{\hbar \to 0}{\to} \; \sum_{k=0}^\infty \frac{\{ \chi , \bullet \}^{k} \left[ \frac{\partial \chi}{\partial E} \right] }{(k+1)!} \,,
\end{aligned} \label{eq:td_quantum_def}
\end{align}
which is \eqref{eq: time delay as energy derivative} with $\w$ substituted by $E$. 
As a last remark, note that \eqref{def: Time delay} or the first line of \eqref{eq:td_quantum_def} are definitions of the time delay where \emph{inelastic} channel (e.g. $2 \to 3$ process) contributions can be computed from the brackets/commutators~\cite{Kim:2025hpn,Alessio:2025flu}.

Similar to the polarisation rotation angle, the time delay is experimentally measured by comparing two different trajectories (impact parameter $b$ and $b_0$) of light/gravitational waves emanating from a common source.\footnote{This setup is similar to the hard IR cut-off considered in ref.~\cite{Bucciotti:2026svg}.} 
When the source ``blinks,'' the time delay is measured as the time difference between observed ``blinks'' of the two trajectories. 
In 4d, the adopted definition of the time delay \eqref{def: Time delay} computes the (divergent) limit $b_0 \to \infty$. 
The notation $\D t (b;b_0)$ will be used for the time delay when referring to its operational definition (how it is measured experimentally). 

Causality is understood to imply the ``positivity'' of the time delay~\cite{Wigner:1955zz,Camanho:2014apa,Bonifacio:2017nnt,Hinterbichler:2017qyt,Hinterbichler:2017qcl,AccettulliHuber:2020oou,deRham:2020zyh}. 
We adopt the (infrared) causality condition as the inequality of the time delay~\cite{Chen:2021bvg,deRham:2021bll,deRham:2022hpx}
\begin{align} \label{eq: IR causality}
    \lim_{b_0 \to \infty} \left( \D t (b;b_0) - \D t_{\text{GR}} (b;b_0) \right) \gtrsim - \frac{1}{\w} \,,
\end{align}
where $\D t_{\text{GR}}$ is the time delay without EFT corrections to general relativity and the lower bound is given by unresolvable negativity. 
This difference is finite in the limit $b_0 \to \infty$, as the divergence in the Shapiro time delay is canceled by the difference.

\section{Effective field theory and computation setup} \label{sec:setup}

\subsection{Effective field theory action}
We consider $FFR + R^3 + R^4$ corrections to Einstein--Maxwell action considered as an EFT with a UV cutoff $\L$. 
The action is given as ($\k := \sqrt{32\pi G}$)
\begin{align}\label{The EFT}\begin{split}
    S=-\int_{1/\L} d^4x\sqrt{-g}\biggr[&\frac{1}{4}F_{\m\n}F^{\m\n}+\frac{\b}{8}F^{\m\n}F^{\r\s}R_{\m\n\r\s}
    \\&+\frac{2}{\k^2}\bigr(R+\frac{\a'^2}{48}(I_1+2G_3)-\z_1\CC^2-\z_2\CC\tilde{\CC}-\z_3\tilde{\CC}^2\bigr)\biggr]\;.
\end{split}\end{align}
To make meaningful statements on causality constraints, we will further assume that the UV cutoff is parametrically lower than the Planck scale: $\L \ll \k^{-1}$~\cite{Camanho:2014apa}. 
The curvature terms appearing in \eqref{The EFT} are defined as
\begin{align}
    I_1 := {R^{\a\b}}_{\m\n}{R^{\m\n}}_{\r\s}{R^{\r\s}}_{\a\b}\;,\quad G_3 := I_1-2{R^{\m\n\a}}_\n{R^{\b\g}}_{\n\s}{R^\s}_{\m\g\a} \,,
\end{align}
for $R^3$ corrections and
\begin{align}
    \mathcal{C} := R_{\m\n\r\s}R^{\m\n\r\s}\;,\quad \tilde{\mathcal{C}} := \frac{1}{2}R_{\m\n\a\b}{\e^{\a\b}}_{\g\d}R^{\g\d\m\n} \,,
\end{align}
for $R^4$ corrections. 
We also use dimensionless Wilson coefficients $C_\#$ and UV cutoff \emph{length} scale $L = \frac{\hbar}{\L}$ to parametrise the EFT corrections of \eqref{The EFT},
\begin{align}
    \b := \frac{C_\b}{\L^2} = \frac{L^2}{\hbar^2} C_\b \;,\quad \a'^2 := \frac{C_{\a'^2}}{\L^4} = \frac{L^4}{\hbar^4} C_{\a'^2} \;,\quad \z_{i=1,2,3} := \frac{C_{\z_i}}{\L^6} = \frac{L^6}{\hbar^6} C_{\z_i} \;, \label{eq:dimlessWC}
\end{align}
which simplifies causality analysis of the theory.\footnote{We use natural units $\hbar=1$ but restore powers of $\hbar$ when $\hbar$ counting becomes relevant.} We also define reparametrised couplings
\begin{align}
    \tilde{\z} := 4(\z_1+\z_3) \,,\quad \z^\pm := 4(\z_1-\z_3\pm\frac{i}{2}\z_2) \,,\quad \z := 4(\z_1-\z_3) \,,
\end{align}
and similarly their associated Wilson coefficients ($C_{\tilde{\z}}$, $C_{\z^\pm}$, and $C_{\z}$). 
The EFT corrections in the observables will be considered to leading order in the couplings.

The considered class of theories is motivated by string theory, where the EFT action \eqref{The EFT} arises as an IR EFT of various string theories~\cite{AccettulliHuber:2020oou}. 
The $R^3$-type correction of the form $I_1 + G_3$ is known to arise in the IR limit of bosonic string theory~\cite{Metsaev:1986yb,Brandhuber:2019qpg}. 
Although we allow arbitrary real values of $C_{\a'^2}$ (i.e. we allow \emph{negative} values of $C_{\a'^2}$), $\a'^2 > 0$ is required if the UV completion of the EFT action \eqref{The EFT} is bosonic string theory.
The $R^4$-type corrections of \eqref{The EFT} arise in the IR limit of type-II string theories~\cite{Gross:1986iv}.
Various string theories are known to generate $FFR$ corrections in the IR~\cite{Stieberger:2009hq,Drummond:1979pp,Goon:2016une,Berends:1975ah,Kawai:1985xq}.

For the wavenumber kick and gravitational Faraday rotation, the modifications from beyond-general-relativity terms in \eqref{The EFT} will turn out to be (mostly) \emph{subleading} in $b^{-1}$ counting compared to quantum corrections from general relativity, supporting the claim that quantised general relativity is a consistent EFT that can be used to make universal predictions of (long-distance effects in) quantum gravity~\cite{Donoghue:1994dn}. However, the parametric hierarchy $\L \ll \k^{-1}$ can dominate over the faster $b^{-1}$ fall-off of EFT corrections, in which case the EFT correction effects ($\propto (\frac{L}{b})^{n \ge 4}$) will dominate over quantum gravity corrections from virtual graviton loops ($\propto \frac{G \hbar}{b^2}$) for moderate values of $b$.

\subsection{The geometric optics regime and expansion variables} \label{sec:geoopt}

We study wave (electromagnetic or gravitational) scattering on black holes as a $2 \to 2$ black-hole--photon/graviton scattering problem. 
There are four length scales in the problem: the Compton wavelength of the black hole $\l_C\equiv \hbar m^{-1}$, the wavelength of the massless particle $\l\equiv\w^{-1}$, the spin length of the black hole $a\equiv |\vec{S}|m^{-1}$, and the impact parameter $b\sim1 /|\vec{q}|$. The (classical) geometric optics regime is given by the hierarchy~\cite{Chen:2022clh}
\begin{align}
    \l_C\ll \l\ll a\ll b\;.
\end{align}
The hierarchy can be further refined by the cosmic censorship condition $a \le Gm$ (no naked singularities) and no absorption condition $b \ge b_c$, where $b_c$ is the critical impact parameter corresponding to marginally plunging trajectories~\cite{Hsiao:2019ohy,Bardeen:1973tla,Perlick:2021aok}.\footnote{\label{fn:crit_b}The value of $b_c$ depends on various parameters such as the mass and spin of the black hole, the angular profile of the trajectory, and the helicity of the test particle~\cite{Almeida:2026ewq}. The smallest/largest value of $b_c$ for an extremal Kerr black hole is $b_{c,\rm min} = 2Gm$ / $b_{c,\rm max} = 7Gm$~\cite{Iyer:2009wa}. Since we are considering black holes in effective field theories of gravity, we will simply set $b_c \sim Gm/\e_{\text{UV}}$ where $\e_{\text{UV}}$ is a $\CO(1)$ number.}
Moreover, the EFT approach demands that the energy of massless quanta should not exceed the UV cutoff, $\hbar \w \lesssim \L$, which becomes the condition $L \lesssim \l$ in dimensions of length. 
Incorporating these conditions, the hierarchy can be refined to
\begin{align} \label{Hierarchy of parameters}
    \l_C \ll L \lesssim \l \ll a\le Gm<b_c \le b\;.
\end{align}
This hierarchy of scales is compatible with the eikonal regime of scattering amplitudes. 

We focus on the leading order effects in the $\hbar$ expansion. Since the Compton wavelength $\l_C = \frac{\hbar}{m}$ scales as $\hbar$, we do not consider expansion in $\l_C$. We can identify the following (classical) expansion parameters from the hierarchy \eqref{Hierarchy of parameters},
\begin{align}
    \e_{\text{EFT}} = \frac{L}{b} \;\; \lesssim \;\; \e_{\text{wo}} = \frac{\l}{b} = \frac{1}{\w b} \;\; \ll \;\; \e_{\text{spin}} = \frac{a}{b} \;\; \le \;\; \e_{\text{PM}} = \frac{Gm}{b} \;\; ( \lesssim \e_{\text{UV}} \sim \CO(1) ) \,, \label{eq:eps_hier}
\end{align}
where $\e_{\text{EFT}}$ parametrises the EFT corrections, $\e_{\text{wo}}$ parametrises the wave-optics corrections to geometric optics, $\e_{\text{spin}}$ parametrises the black hole spin corrections, and $\e_{\text{PM}}$ parametrises the post-Minkowskian corrections. 
The $\CO (1)$ upper bound $\e_{\text{UV}}$ has been set to prevent the impact parameter $b$ becoming smaller than the critical impact parameter $b_c$. 
The hierarchy reveals that wave-optics corrections are more important than EFT corrections ($\e_{\text{EFT}} \lesssim \e_{\text{wo}}$), therefore polarisation of the massless particles must be taken into account when analysing effects of beyond-general-relativity corrections.\footnote{The EFT condition $\w \ll \L^2 r$ presented as eq.(5.15) in ref.~\cite{deRham:2020zyh} translates to $\e_{\text{EFT}}^2 \ll \e_{\text{wo}}$ in our notation, which is weaker than the conditions of \eqref{eq:eps_hier}.}

The higher spin-induced multipole moments\footnote{Quadrupole moments or higher.} of spinning black holes in EFTs of gravity are known to differ from that of Kerr black holes in general relativity~\cite{Cano:2022wwo}. We limit our analysis to linear-in-spin effects to avoid complications related to higher multipole moments.

\subsection{Computation setup} \label{sec:kinematics}
\subsubsection{Impact parameter space Fourier transform}

Our analyses are based on $2 \to 2$ massive-massless eikonal amplitudes $i M_{\text{eik}} = e^{i \delta} - 1$, where the massive particle represents a spinning black hole and the massless particle represents the incident electromagnetic or gravitational wave. The eikonal phase (matrix) $\delta$ is interpreted as the Magnusian (matrix) for computing observables; see sec.~\ref{sec:Magnusian}.

\begin{figure}[h]
    \centering
    \begin{tikzpicture}
        \node[inner sep=0] at (0,0){
            \begin{fmffile}{compton}
                \parbox{100pt}{
                    \begin{fmfgraph*}(100,100)
                        \fmfstraight
                        \fmfleft{i1,c1,o1}
                        \fmfright{i2,c2,o2}
                        \fmf{phantom}{c1,e1,c2}
                         \fmffreeze
                        \fmf{plain,width=3}{i1,e1,o1}
                        \fmf{wiggly}{i2,e1,o2}     
                        \fmfv{decor.shape=circle,decor.filled=empty,decor.size=50,label=\Large\(M_4\),label.dist=0}{e1}
                        \fmfv{label=$h_2$,label.angle=0}{i2}
                        \fmfv{label=$-h_3$,label.angle=0}{o2}
                    \end{fmfgraph*}
                }
            \end{fmffile}
        };
        \draw[->] (-46pt,40pt) -- (-26pt,20pt);
        \node at (-45pt, 21pt) {\(-p_4^\mu\)};
        \draw[->] (-46pt,-40pt) -- (-26pt,-20pt);
        \node at (-39pt,-21pt) {\(p_1^\mu\)};
        \draw[->] (50pt,40pt) -- (30pt,20pt);
        \node at (47pt,21pt) {\(-k_3^\mu\)};
        \draw[->] (50pt,-40pt) -- (30pt,-20pt);
        \node at (47pt,-21pt) {\(k_2^\mu\)};
        \draw[->] (-12pt, 29pt) -- (12pt,29pt);
        \node at (0pt, 37pt) {\(q^\mu\)};
    \end{tikzpicture}
    \caption{Kinematics of the Compton amplitude $M_4$ in the all-incoming convention. Bold solid lines denote massive particles and wavy lines denote massless particles, whose helicity configurations are $h_2$ and $-h_3$.
    }
    \label{fig : Momenta assignment}
\end{figure}
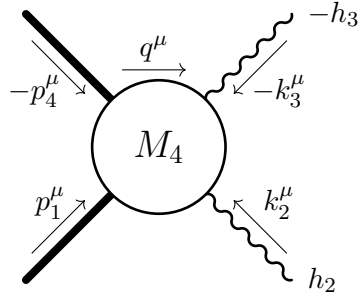

The Magnusian matrix is computed by Fourier-transforming the Compton amplitude (fig.~\ref{fig : Momenta assignment}) to impact parameter space.
We assign $p_1^\m$ ($p_4^\m$) as the incoming (outgoing) momentum of the massive particle representing the spinning black hole (of mass $m$) and $k_2^\m$ ($k_3^\m$) as the incoming (outgoing) wavenumber of the massless particle representing light and gravitational waves with helicity $h_2$ ($h_3$). 
All external momenta satisfy on-shell conditions ($p_1^2 = p_4^2 = m^2$ and $k_2^2 = k_3^2 = 0$) and the transfer momentum is defined as $q^\m = p_1^\m - p_4^\m = k_3^\m - k_2^\m$.
The spin four-vector (Pauli-Lubanski pseudovector) of the massive particle is denoted as $S^\m$, which satisfies the orthogonality condition $p_1 \cdot S = 0$.
We identify the rest frame of $p_1^\m$ as the rest frame of the black hole background and define $\w := \frac{p_1 \cdot k_2}{m}$ as the frequency of the incoming light/gravitational wave, setting the positive-$z$ direction as the incoming direction.

The Fourier transform to impact parameter space is given as
\begin{align}
    \exp\left( i \chi^{(h_{ i},-h_{f})}(\vec{b}) \right) - 1 = \int\frac{d^D q}{(2\pi)^{D-2}} \delta (2 p_1 \cdot q) \delta (2 k_2 \cdot q)  e^{- i q \cdot b} \; i M^{(h_{i},-h_{f})}_4 \,, \label{eq:IPSdef}
\end{align}
where it is understood that we only keep the leading terms in the $\hbar$ expansion. We collect the matrix elements $\chi_{IJ} = \chi^{(J,-I)}$ into the Magnusian matrix $\chi$ as\footnote{The sign difference is due to the all-incoming convention of Compton amplitudes.}
\begin{align}
    \chi & =
    \begin{pmatrix}
        \chi^{(+-)} & \chi^{(++)} \\
        \chi^{(--)} & \chi^{(-+)}
    \end{pmatrix} \,. \label{eq:MagMatExpDef}
\end{align}
We will find that off-diagonal terms of the Magnusian matrix are non-vanishing, which implies that circular and linear polarisations will become elliptical after scattering. The following $2 \times 2$ matrices will be used to expand the Magnusian matrix.
\begin{align}
    I_{2 \times 2} = \begin{pmatrix}
        1&0\\0&1
    \end{pmatrix} \,,\; 
    \s_1 = \begin{pmatrix}
        0&1\\1&0
    \end{pmatrix} \,,\;  
    \s_2 = \begin{pmatrix}
        0& -i\\i &0
    \end{pmatrix} \,,\;  
    \s_3 = \begin{pmatrix}
        1&0\\0&-1
    \end{pmatrix} \,,\; 
    \s_\pm = \frac{\s_1 \pm i \s_2}{2} \,.
\end{align}

Although correctly defining the impact parameter space can be subtle, the subtleties only arise at higher orders in the loop or $\hbar$ expansions and therefore can be neglected~\cite{Chen:2022clh,DiVecchia:2023frv}.\footnote{The subtleties related to obtaining observables from spinning eikonal amplitudes (such as translation operators considered in ref.~\cite{Luna:2023uwd}) are taken care of by the Magnusian approach~\cite{Kim:2025hpn}; also see footnote \ref{fn:eik_Mag}.} 
We adopt $\hbar$ counting used in the post-Minkowskian dynamics literature~\cite{Kosower:2018adc}
\begin{align}
    m \to m\;,\quad p^\m \to p^\m \;,\quad k^\m \to \hbar k^\m\;,\quad S^\m \to \frac{S^\m}{\hbar}\;,\quad G \to \frac{G}{\hbar}\;,\quad C_\# \to C_\# \;,
\end{align}
where $p^\m$ stands for any massive momenta and $k^\m$ stands for any massless momenta, including the transfer momentum $q^\m$. Introducing the dimensionless Wilson coefficients $C_\#$ as in \eqref{eq:dimlessWC} takes care of the EFT Wilson coefficient $\hbar$ counting. The $\e$ hierarchy \eqref{eq:eps_hier} in momentum space can be written as
\begin{align}
    \e_{\text{EFT}} \sim L q \;\; \lesssim \;\; \e_{\text{wo}} \sim \frac{q}{\w} \;\; \ll \;\; \e_{\text{spin}} \sim qa \;\; \le \;\; \e_{\text{PM}} \sim Gmq \;\; \lesssim \;\; \e_{\text{UV}}\,, \label{eq:eps_hier2}
\end{align}
where $q = |\vec{q}|$ is the size of the transfer momentum. Combining dimensional analysis with analyticity of the master integral coefficients in Mandelstam invariants, it can be shown that $\e_{\text{wo}}$ corrections are absent at tree order and odd powers of $\e_{\text{wo}}$ corrections must accompany spin dependence~\cite{Chen:2022clh}.

\begin{figure}[h]
    \centering
    \includegraphics[width=0.5\textwidth]{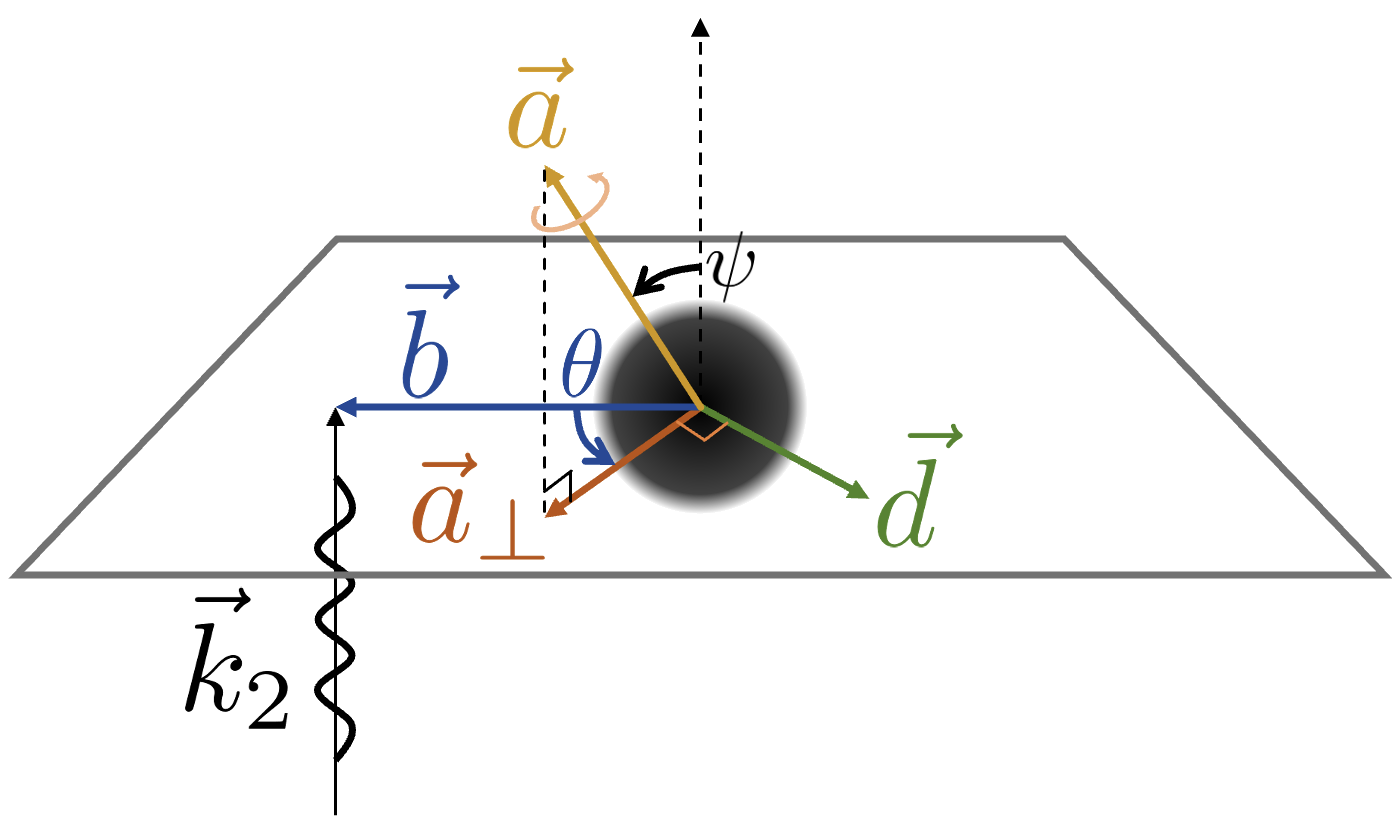}
    \caption{Configuration of spin variables in impact parameter space. $\vec{a} := \vec{S}/m$ is the spin-length vector of the black hole, $\vec{a}_\bot$ is the projection of $\vec{a}$ onto the $\vec{b}$-plane, and $\vec{d} := \hat{k}_2\times \vec{a}_\bot$ is the vector in the $\vec{b}$-plane orthogonal to $\vec{a}_\perp$.
    }
    \label{fig:kinematics}
\end{figure}

We fix the Compton amplitude's frame as
\begin{align}
\begin{aligned}
    p_1^\m &= \left(\sqrt{m^2+|\vec{k}_2|^2} \,,\; 0\,,\; 0 \,,\; -|\vec{k}_2|\right) \,,
    \\ k_2^\m &= \left(|\vec{k}_2| \,,\; 0\,,\; 0 \,,\;|\vec{k}_2| \right) \,,
\end{aligned}
\end{align}
and Fourier-transform in the $xy$-plane of $q^\m$ to go to impact parameter space, i.e. 
\begin{align}
    \exp \left( i \chi^{(h_{ i},-h_{f}) }(\vec{b}) \right)-1 = \frac{i}{4m\w} \int \frac{dq_x dq_y}{(2\pi)^2} \; e^{i\vec{q}\cdot\vec{b}} \; M^{(h_{i},-h_{f})}_4 \,. \label{eq:IPStrsfDef1}
\end{align}
We define the orthogonal vector $n_\m = \e_{\m\a\b\g} p_{1}^{\a} k_{2}^{\b} q^{\g}$ and adopt spinor conventions of ref.~\cite{Chung:2018kqs}. The definition of spin vector variables ($\vec{a}$, $\vec{a}_\perp$, and $\vec{d}$) are shown in fig.~\ref{fig:kinematics}.
We also define holomorphic variables~\cite{AccettulliHuber:2020oou}
\begin{align}
   \fb := \frac{b_x + ib_y}{2}\;,\;\bfb := \frac{b_x - ib_y}{2}\text{ and similarly }\fd\;,\;\bfd\;,\;\fa\;,\;\bfa\;,
\end{align}
because holomorphic transfer momenta, $\bfq := q_x - iq_y$ and $\fq := q_x + iq_y$, appear in spinor brackets; see app.~\ref{app:spinors}. Note that the combination $\vec{q}\cdot\vec{b}=\fq\bfb+\bfq\fb$ appears in the exponent when evaluating the Fourier transform \eqref{eq:IPStrsfDef1}. 

When analysing observables, we choose coordinates where the impact parameter lies along the positive $x$-axis: $\vec{b} = b \hat{x}$. The spin vector $\vec{S}$ is parametrised by two angles $\psi$ and $\th$,
\begin{align}
    \vec{S} = ma(\sin\psi\cos\th,\sin\psi\sin\th,\cos\psi) \;,
\end{align}
as shown in fig.~\ref{fig:kinematics}. We list some relevant inner products formed by the spin vector: 
\begin{align}
    \vec{k}_2\cdot\vec{S}=m\w a\cos\psi\;,\quad \vec{a}_\bot\cdot\vec{b}=ab\sin\psi\cos\th\;,\quad \vec{d}\cdot\vec{b}=-ab\sin\psi\sin\th\;.
\end{align}

\subsubsection{Cut-construction of tree-level and one-loop Compton amplitudes}

The nonanalyticity of the transfer momentum $\vec{q}$ near $\vec{q} = 0$ is the most important feature of the Compton amplitude when Fourier-transforming to impact parameter space, since the nonanalyticity determines the large-$\vec{b}$ behaviour which corresponds to the eikonal regime. This regime is governed by long-distance exchange of virtual gravitons and is encoded in the cut-constructible part of the Compton amplitude~\cite{Bjerrum-Bohr:2013bxa,Neill:2013wsa}. We provide a brief summary of how the cut-constructible part of the Compton amplitudes can be obtained. The seed amplitudes are summarised in app.~\ref{app:seed_amps}.

\begin{figure}[h]
    \centering
    \begin{tikzpicture}
        \node[inner sep=0] at (0,0){
            \begin{fmffile}{tree}
                \parbox{150pt}{
                    \begin{fmfgraph*}(150,100)
                        \fmfstraight
                        \fmfleft{i1,c1,o1}
                        \fmfright{i2,c2,o2}
                        \fmf{phantom}{c1,b1,e1,e2,e3,b2,c2}
                         \fmffreeze
                        \fmf{plain,width=3}{i1,b1,o1}
                        \fmf{dbl_wiggly}{b1,b2}
                        \fmf{wiggly}{i2,b2,o2}
                        \fmfv{decor.shape=circle,decor.filled=shaded,decor.size=40}{b1}
                        \fmfv{decor.shape=circle,decor.filled=empty,decor.size=40,label=I,label.dist=0}{b2}
                    \end{fmfgraph*}
                }
            \end{fmffile}
        };
        \draw[->] (-75pt,40pt) -- (-65pt,20pt);
        \node at (-85pt, 27pt) {\(-p_4^\mu\)};
        \draw[->] (-75pt,-40pt) -- (-65pt,-20pt);
        \node at (-80pt,-27pt) {\(p_1^\mu\)};
        \draw[->] (79pt,40pt) -- (69pt,20pt);
        \node at (90pt,27pt) {\(-k_3^\mu\)};
        \draw[->] (79pt,-40pt) -- (69pt,-20pt);
        \node at (86pt,-27pt) {\(k_2^\mu\)};
        \draw[->] (-15pt,10pt) -- (15pt,10pt);
        \node at (0pt,17pt) {\(q^\mu\)};
            \draw[draw=white,line width=6pt] (0pt,7pt) -- (0pt,-7pt);
            \draw[draw=black, dashed,line width=1pt] (0pt,7pt) -- (0pt,-7pt);
            \node[font=\normalsize, rotate=90] at (0pt, -9pt) {\ScissorRightBrokenBottom};
    \end{tikzpicture}
    \caption{$t$-channel gluing for the tree-level Compton amplitude. The left shaded blob denotes BH--BH--graviton 3pt amplitude and the right blob denotes the 3pt amplitude corresponding to the interaction ${\rm I}\in\{{\rm EH(M)},FFR,R^3,R^4\}$.
    }
    \label{fig:t_cut}
\end{figure}
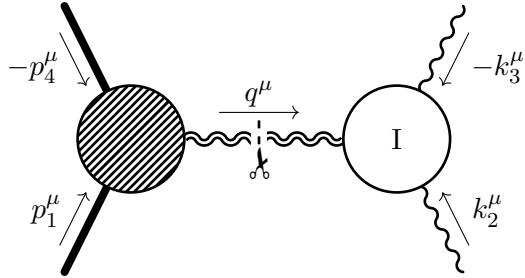

For tree-level Compton amplitudes, the nonanalytic behaviour near $\vec{q} = 0$ is captured by the $t$-channel factorisation pole. The residue of the $t$-channel pole can be constructed by ``gluing'' two on-shell 3-point amplitudes as shown in fig.~\ref{fig:t_cut}. We build the relevant part of the tree-level Compton amplitude by gluing BH(black hole)-BH-graviton 3pt amplitude and the 3pt amplitude corresponding to Einstein(--Hilbert)--Maxwell theory (EHM or EH depending on context) or EFT corrections ($FFR$, $R^3$, and $R^4$) considered in \eqref{The EFT}. 

\begin{figure}[h]
    \centering
    \begin{tikzpicture}
        \node[inner sep=0] at (0,0){
            \begin{fmffile}{one_loop}
                \parbox{150pt}{
                    \begin{fmfgraph*}(150,100)
                        \fmfstraight
                        \fmfleft{i1,c1,o1}
                        \fmfright{i2,c2,o2}
                        \fmf{phantom}{c1,b1,e1,e2,e3,b2,c2}
                         \fmffreeze
                        \fmf{plain,width=3}{i1,b1,o1}
                        \fmf{dbl_wiggly,left=0.6}{b1,b2}
                        \fmf{dbl_wiggly,right=0.6}{b1,b2}
                        \fmf{wiggly}{i2,b2,o2}
                        \fmfv{decor.shape=circle,decor.filled=shaded,decor.size=40}{b1}
                        \fmfv{decor.shape=circle,decor.filled=empty,decor.size=40,label=I,label.dist=0}{b2}
                    \end{fmfgraph*}
                }
            \end{fmffile}
        };
        \draw[->] (-75pt,40pt) -- (-65pt,20pt);
        \node at (-85pt, 27pt) {\(-p_4^\mu\)};
        \draw[->] (-75pt,-40pt) -- (-65pt,-20pt);
        \node at (-80pt,-27pt) {\(p_1^\mu\)};
        \draw[->] (79pt,40pt) -- (69pt,20pt);
        \node at (90pt,27pt) {\(-k_3^\mu\)};
        \draw[->] (79pt,-40pt) -- (69pt,-20pt);
        \node at (86pt,-27pt) {\(k_2^\mu\)};
        \draw[->] (-15pt,38pt) -- (15pt,38pt);
        \node at (3pt,47pt) {\(l^\mu + q^\mu\)};
        \draw[->] (15pt,-38pt) -- (-15pt,-38pt);
        \node at (0pt,-47pt) {\(l^\mu\)};
            \draw[draw=white,line width=6pt] (0pt,37pt) -- (0pt,23pt);
            \draw[draw=black, dashed,line width=1pt] (0pt,37pt) -- (0pt,23pt);
            \node[font=\normalsize, rotate=90] at (0pt, 21pt) {\ScissorRightBrokenBottom};
            \draw[draw=white,line width=6pt] (0pt,-23pt) -- (0pt,-37pt);
            \draw[draw=black, dashed,line width=1pt] (0pt,-23pt) -- (0pt,-37pt);
            \node[font=\normalsize, rotate=-90] at (0pt, -21pt) {\ScissorRightBrokenBottom};
    \end{tikzpicture}
    \caption{$t$-channel two-graviton unitarity cut for one-loop Compton amplitude. The left shaded blob denotes BH--BH--two-graviton 4pt amplitude and the right blob denotes the 4pt amplitude corresponding to the interaction ${\rm I}\in\{{\rm EH(M)},FFR,R^3,R^4\}$.
    }
    \label{fig:t_2h_cut}
\end{figure}
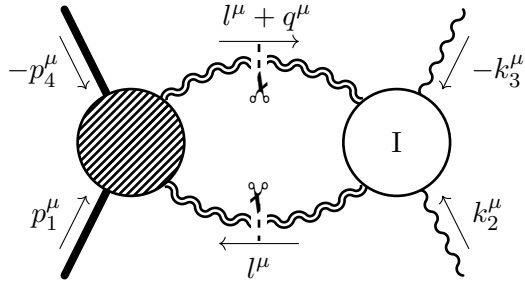

\begin{figure}[h]
    \centering
    \begin{minipage}[b]{0.3\textwidth}
        \centering
        \begin{tikzpicture}
            \node[inner sep=0] at (0,0){
                \begin{fmffile}{master1}
                    \parbox{120pt}{
                            \begin{fmfgraph*}(120,60)
                                \fmfstraight
                                \fmfleft{i1,o1}
                                \fmfright{i2,o2}
                                \fmf{phantom}{i1,v1,v2,v3,i2}
                                \fmf{phantom}{o1,u1,u2,u3,o2}
                                 \fmffreeze
                                \fmf{plain,width=3}{i1,i2}
                                \fmf{wiggly}{o1,u2,o2}
                                \fmf{wiggly}{u2,v1}
                                \fmf{wiggly}{v3,u2}
                            \end{fmfgraph*}
                        }
                \end{fmffile}
                };
        \node[] at (0pt,-45pt) {\(\mathcal{I}_{\Delta}\)};
        \end{tikzpicture}
    \end{minipage}
    \hfill
    \begin{minipage}[b]{0.3\textwidth}
        \centering
        \begin{tikzpicture}
            \node[inner sep=0] at (0,0){
                \begin{fmffile}{master2}
                    \parbox{120pt}{
                            \begin{fmfgraph*}(120,60)
                                \fmfstraight
                                \fmfleft{i1,o1}
                                \fmfright{i2,o2}
                                \fmf{phantom}{i1,v1,v2,v3,i2}
                                \fmf{phantom}{o1,u1,u2,u3,o2}
                                 \fmffreeze
                                \fmf{plain,width=3}{i1,i2}
                                \fmf{wiggly}{o1,u1,u3,o2}
                                \fmf{wiggly}{v1,u1}
                                \fmf{wiggly}{u3,v3}
                            \end{fmfgraph*}
                        }
                \end{fmffile}
                };
        \node[] at (0pt,-45pt) {\(\mathcal{I}_{\square}\)};
        \end{tikzpicture}
    \end{minipage}
    \hfill
    \begin{minipage}[b]{0.3\textwidth}
        \centering
        \begin{tikzpicture}
            \node[inner sep=0] at (0,0){
                \begin{fmffile}{master3}
                    \parbox{120pt}{
                            \begin{fmfgraph*}(120,60)
                                \fmfstraight
                                \fmfleft{i1,o1}
                                \fmfright{i2,o2}
                                \fmf{phantom}{i1,v1,v2,v3,i2}
                                \fmf{phantom}{o1,u1,u2,u3,o2}
                                 \fmffreeze
                                \fmf{plain,width=3}{i1,i2}
                                \fmf{wiggly}{o1,u1,u3,o2}
                                \fmf{wiggly}{v1,u3}
                                \fmf{wiggly,rubout=3.5}{u1,v3}
                            \end{fmfgraph*}
                        }
                \end{fmffile}
                };
        \node at (0pt,-45pt) {\(\mathcal{I}_{\times}\)};
        \end{tikzpicture}
    \end{minipage}
    \caption{The relevant master integrals: massive triangle, box, and crossed-box. Bold solid lines are massive propagators and wavy lines are massless propagators}
     \label{fig:MIs}
\end{figure}
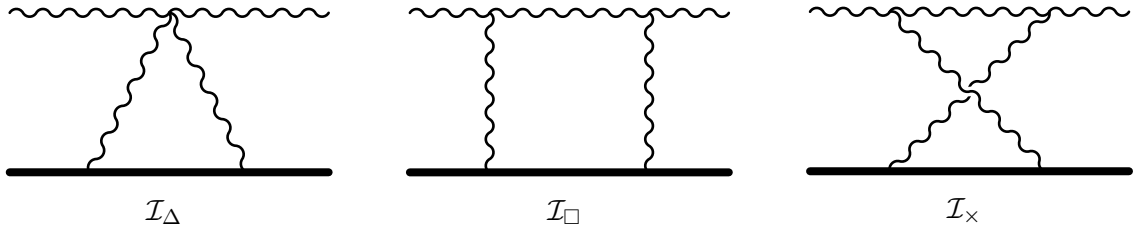

At one-loop level, the relevant nonanalytic behaviour is captured by the two-graviton exchange in the $t$-channel. Similar to the tree case, the relevant loop integrand can be computed by unitarity methods~\cite{Bern:1994zx,Bern:1994cg,Bern:1997sc} where we glue two 4-point amplitudes across the $t$-channel two-graviton cut as shown in fig.~\ref{fig:t_2h_cut}. The master integral coefficients are evaluated from the cut-constructed loop integrands using two different methods for cross-validation: through Forde's method~\cite{Forde:2007mi} and through the integration-by-parts (IBP) package \texttt{LiteRed2}~\cite{Lee:2012cn,Lee:2013mka}. We use momentum parametrisation of ref.~\cite{Chen:2022clh} when using Forde's method to compute the master integral coefficients. Because we only consider leading $\hbar$ effects, the relevant master integrals are the massive triangle ($\CI_{\D}$), the box ($\CI_\square$), and the crossed-box ($\CI_{\times}$) integrals schematically given in fig.~\ref{fig:MIs}.

\section{Faraday rotation and spin Hall effect in EFTs of gravity} \label{sec:observables1}

We compute the Magnusian matrix at one-loop order to leading order in the EFT couplings and use it to study gravitational Faraday rotation and gravitational spin Hall effect for electromagnetic and gravitational waves in EFTs of gravity. To avoid clutter, we describe their computation schematically as diagrams similar to figs.~\ref{fig:t_cut}-\ref{fig:t_2h_cut}.

Since wave-optics corrections ($\e_{\text{wo}} = (\w b)^{-1}$) are only relevant at one-loop order~\cite{Chen:2022clh}, we present tree-level Magnusian without wave-optics corrections and present one-loop Magnusian with wave-optics corrections to leading order for each interaction.\footnote{To order $\CO (\e_{\text{wo}}^1)$ if linear order corrections are present, otherwise to order $\CO (\e_{\text{wo}}^2)$.} The helicity Compton amplitudes are organised in a matrix form similar to the Magnusian matrix \eqref{eq:MagMatExpDef}. When computing observables, we only consider wave-optics corrections to linear order in the Magnusian.

\subsection{Photon scattering} \label{sec:phosc}

The tree-level Compton amplitude is computed as a sum of two contributions,
\begin{align}
\begin{aligned}
    \left. M_4^{\g} \right|_{G^1} &\sim
    \quad
\begin{fmffile}{photon_LO_EHM}
\parbox{80pt}{
    \begin{fmfgraph*}(80,60)
        \fmfstraight
        \fmfleft{i1,c1,o1}
        \fmfright{i2,c2,o2}
        \fmf{phantom}{c1,b1,e1,e2,e3,b2,c2}
         \fmffreeze
        \fmf{plain,width=2}{i1,b1,o1}
        \fmf{dbl_wiggly}{b1,b2}
        \fmf{wiggly}{i2,b2,o2}
        \fmfv{decor.shape=circle,decor.filled=shaded,decor.size=30}{b1}
        \fmfv{decor.shape=circle,decor.filled=empty,decor.size=30,label=EHM,label.dist=0}{b2}
        \fmf{phantom}{i1,d1,i2}
        \fmf{phantom}{o1,d2,o2}
        \fmf{dashes,tension=2,rubout=5}{d1,d2}
    \end{fmfgraph*}
}
\end{fmffile}
    \quad + \quad 
\begin{fmffile}{photon_LO_FFR}
\parbox{80pt}{
    \begin{fmfgraph*}(80,60)
        \fmfstraight
        \fmfleft{i1,c1,o1}
        \fmfright{i2,c2,o2}
        \fmf{phantom}{c1,b1,e1,e2,e3,b2,c2}
         \fmffreeze
        \fmf{plain,width=2}{i1,b1,o1}
        \fmf{dbl_wiggly}{b1,b2}
        \fmf{wiggly}{i2,b2,o2}
        \fmfv{decor.shape=circle,decor.filled=shaded,decor.size=30}{b1}
        \fmfv{decor.shape=circle,decor.filled=empty,decor.size=30,label=$FFR$,label.dist=0}{b2}
        \fmf{phantom}{i1,d1,i2}
        \fmf{phantom}{o1,d2,o2}
        \fmf{dashes,tension=2,rubout=5}{d1,d2}
    \end{fmfgraph*}
}
\end{fmffile}
    \\ & = \left(\frac{\k}{2}\right)^2 \frac{4m^2\w^2}{|\vec{q}|^2}\left( 1-\frac{i(n\cdot S)}{m^2\w} \right) \left[ I_{2\times 2}-\frac{\b}{4}\begin{pmatrix}
        0&\fq^2\\\bfq^2&0
    \end{pmatrix}\right]\;.
\end{aligned}
\end{align}
The holomorphic transfer momentum dependence (dependence on $\texttt{q}$ and $\bar{\texttt{q}}$) should be viewed as EFT corrections ($\e_{\text{EFT}} = L/b$) rather than wave-optics corrections ($\e_{\text{wo}} = \l / b$). The corresponding Magnusian matrix (or eikonal phase matrix) is
\begin{align} \label{eq: Photon Eik G1}
    \chi^{\g}_{G^1} = 4m\w G\left[\left(\log\frac{b_0}{b}-\frac{\vec{d}\cdot\vec{b}}{b^2}\right)I_{2\times 2}+\frac{\b}{8\bfb^2}\left(1-\frac{2\bfd}{\bfb}\right)\s_++\frac{\b}{8\fb^2}\left(1-\frac{2\fd}{\fb}\right)\s_-\right]\;.
\end{align}

The one-loop Compton amplitude is computed as a sum of three contributions,
\begin{align}
\begin{aligned}
    \left. M^{\g}_{4} \right|_{G^2}\sim&\quad
    \begin{fmffile}{photon_NLO_EHM}
    \parbox{80pt}{
        \begin{fmfgraph*}(80,60)
            \fmfstraight
            \fmfleft{i1,c1,o1}
            \fmfright{i2,c2,o2}
            \fmf{phantom}{c1,b1,e1,e2,e3,b2,c2}
            \fmf{phantom}{i1,d1,i2}
            \fmf{phantom}{o1,d2,o2}
            \fmf{phantom}{d1,f1,f2,f3,d2}
             \fmffreeze
            \fmf{plain,width=2}{i1,b1,o1}
            \fmf{dbl_wiggly,left=0.7}{b1,b2}
            \fmf{dbl_wiggly,right=0.7}{b1,b2}
            \fmf{wiggly}{i2,b2,o2}
            \fmfv{decor.shape=circle,decor.filled=shaded,decor.size=30}{b1}
            \fmfv{decor.shape=circle,decor.filled=empty,decor.size=30,label=EHM,label.dist=0}{b2}
            \fmf{dashes,rubout=5}{d1,f1}
            \fmf{dashes,rubout=5}{d2,f3}
        \end{fmfgraph*}
    }
    \end{fmffile}
        \quad + \quad
    \begin{fmffile}{photon_NLO_R3}
    \parbox{80pt}{
        \begin{fmfgraph*}(80,60)
            \fmfstraight
            \fmfleft{i1,c1,o1}
            \fmfright{i2,c2,o2}
            \fmf{phantom}{c1,b1,e1,e2,e3,b2,c2}
            \fmf{phantom}{i1,d1,i2}
            \fmf{phantom}{o1,d2,o2}
            \fmf{phantom}{d1,f1,f2,f3,d2}
             \fmffreeze
            \fmf{plain,width=2}{i1,b1,o1}
            \fmf{dbl_wiggly,left=0.7}{b1,b2}
            \fmf{dbl_wiggly,right=0.7}{b1,b2}
            \fmf{wiggly}{i2,b2,o2}
            \fmfv{decor.shape=circle,decor.filled=shaded,decor.size=30}{b1}
            \fmfv{decor.shape=circle,decor.filled=empty,decor.size=30,label=$R^3$,label.dist=0}{b2}
            \fmf{dashes,rubout=5}{d1,f1}
            \fmf{dashes,rubout=5}{d2,f3}
        \end{fmfgraph*}
    }
    \end{fmffile}\quad+\quad \begin{fmffile}{photon_NLO_FFR}
    \parbox{80pt}{
        \begin{fmfgraph*}(80,60)
            \fmfstraight
            \fmfleft{i1,c1,o1}
            \fmfright{i2,c2,o2}
            \fmf{phantom}{c1,b1,e1,e2,e3,b2,c2}
            \fmf{phantom}{i1,d1,i2}
            \fmf{phantom}{o1,d2,o2}
            \fmf{phantom}{d1,f1,f2,f3,d2}
             \fmffreeze
            \fmf{plain,width=2}{i1,b1,o1}
            \fmf{dbl_wiggly,left=0.7}{b1,b2}
            \fmf{dbl_wiggly,right=0.7}{b1,b2}
            \fmf{wiggly}{i2,b2,o2}
            \fmfv{decor.shape=circle,decor.filled=shaded,decor.size=30}{b1}
            \fmfv{decor.shape=circle,decor.filled=empty,decor.size=30,label=$FFR$,label.dist=0}{b2}
            \fmf{dashes,rubout=5}{d1,f1}
            \fmf{dashes,rubout=5}{d2,f3}
        \end{fmfgraph*}
    }
    \end{fmffile} \quad.
\end{aligned}
\end{align}
The exponentiation of the tree-level eikonal phase reproduces the box and the crossed-box contributions, therefore only the triangle contribution contributes to the eikonal phase (interpreted as the Magnusian). The Magnusian matrix is
\begin{align} \label{eq: Photon Eik G2}
\begin{split}
    \chi_{G^2}^{\g} &=\frac{15\pi G^2m^2\w}{4b}\left[\left(1-\frac{4(\vec{d}\cdot\vec{b})}{3b^2} 
    \right)I_{2\times 2} 
    - \frac{\hat{k}_2\cdot\vec{a}}{3\w b^2} \, \s_3
    \right.
    \\&\left.\phantom{as\left(\frac{4(\vec{d}\cdot\vec{b})}{3b^2}\right)} - \a'^2\left(\frac{3}{80b^4}+\frac{15}{64\w^2b^6}\right)I_{2\times 2} \right.
    \\&\left.\phantom{as\left(\frac{4(\vec{d}\cdot\vec{b})}{3b^2}\right)}  + \b \left(\frac{3}{16\bfb^2} - \frac{10\fb\bfd+2\bfb\fd}{4b^2\bfb^2} + \frac{1}{\w^2 b^2} \left( \frac{5}{64\bfb^2}+\frac{21\fb\bfd+15\bfb\fd}{8 b^2\bfb^2} \right) \right)\s_+  \right.
    \\&\left.\phantom{as\left(\frac{4(\vec{d}\cdot\vec{b})}{3b^2}\right)} + \b\left(\frac{3}{16\fb^2}-\frac{10\bfb\fd+2\fb\bfd}{4b^2\fb^2}+\frac{1}{\w^2b^2}\left(\frac{5}{64\fb^2}+\frac{21\bfb\fd+15\fb\bfd}{8b^2\fb^2}\right)\right)\s_-\right] \,,
\end{split}
\end{align}
where we have included leading wave-optics corrections ($\e_{\text{wo}} = (\w b)^{-1}$) to all interactions. 

\subsubsection{Gravitational Faraday rotation}

The polarisation rotation angle \eqref{Pol rotation angle} is computed as the $\s_3$ component of the Magnusian matrix, which arises at one-loop order \eqref{eq: Photon Eik G2}:
\begin{align}
    \a_{\g} &= \frac{5\pi G^2m^2 (\hat{k}_2 \cdot \vec{a})}{4b^3} = \frac{5\pi G^2m^2a\cos\psi}{4b^3} \;. \label{eq:photonGFR}
\end{align}
This is the general relativity contribution~\cite{Chen:2022clh}.
An interesting aspect of \eqref{eq:photonGFR} is that the rotation angle does not depend on the energy of the photon; gravitational Faraday rotation of light is a ``geometric'' effect that does not depend on the details of the electromagnetic wave but only depends on the background spacetime generated by the black hole, even in EFTs of gravity. We will see that this is not the case for gravitational waves in sec.~\ref{sec:gravsc}.

\subsubsection{Gravitational spin Hall effect}
The gravitational spin Hall effect (of light) refers to polarisation-dependent scattering of light on curved spacetimes~\cite{Oancea:2019pgm}. In our setup, this effect manifests itself as polarisation-dependent wavenumber kick. The wavenumber kick is computed using \eqref{def: Deflection angle}, 
\begin{align}
\begin{aligned}
    \CK^{i\in\{\para,\bot\}} &= \CK_0^i \, I_{2\times 2} + \sum_{j=1}^3 \CK_j^i \, \s_j
    \\ &= \left( e^{\{\chi,\bullet\}}-1 \right) [k_2^{i}] = \{\chi_{G^1}+\chi_{G^2},k_2^{i}\}+\frac{1}{2}\bigr\{\chi_{G^1},\{\chi_{G^1},k^{i}_2\}\bigr\} + \CO(G^3) \,,
\end{aligned}
\end{align}
where the superscript $\para$($\bot$) indicates the direction parallel(orthogonal) to the impact parameter. Since we set $\vec{b} = b \hat{x}$ and $\vec{k}_2 = k_2 \hat{z}$, $\para$ can be considered as the $x$-component and $\bot$ can be considered as the $y$-component of the wavenumber kick.

The parallel components of the wavenumber kick are
\begin{align}
    \begin{aligned}
        \CK_0^\para&=-\frac{4Gm\w}{b}\left[1+\frac{a\sin\psi\sin\th}{b}+\frac{Gm}{b}\biggr(\frac{15\pi}{16}+\frac{5\pi a\sin\psi\sin\th}{2b}\biggr)-C_{\a'^2}\biggr(\frac{L}{b}\biggr)^4\frac{45\pi Gm}{256b}\right] \,,
        \\ \CK_1^\para&=-\frac{4Gm\w}{b} C_\b\biggr(\frac{L}{b}\biggr)^2\left[1+\frac{3a\sin\psi\sin\th}{b}+\frac{Gm}{b}\biggr(\frac{135\pi}{64}+\frac{45\pi a\sin\psi\sin\th}{4b}\biggr)\right] \,,
        \\ \CK_2^\para&=\frac{4Gm\w}{b} C_\b\biggr(\frac{L}{b}\biggr)^2\left[\frac{3a\sin\psi\cos\th}{b}+\frac{Gm}{b}\biggr(\frac{8a\cos\psi}{b}+\frac{15\pi a\sin\psi\cos\th}{2b}\biggr)\right] \,,
        \\ \CK_3^\para &= \frac{4Gm\w}{b}\times\frac{Gm}{b}\times\frac{1}{\w b}\times\frac{15a\cos\psi}{16b}\,,
    \end{aligned} \label{eq:pho_wk_para}
\end{align}
and the orthogonal components are 
\begin{align}
    \begin{aligned}
        \CK_0^\bot&=-\frac{4Gm\w}{b}\left[\frac{a\sin\psi\cos\th}{b}+\frac{Gm}{b}\biggr(\frac{4a\cos\psi}{b}+\frac{5\pi a\sin\psi\cos\th}{4b}\biggr)\right] \,,
        \\ \CK_1^\bot&=-\frac{4Gm\w}{b} C_\b\biggr(\frac{L}{b}\biggr)^2\left[\frac{3a\sin\psi\cos\th}{b}+\frac{Gm}{b}\biggr(\frac{16a\cos\psi}{b}+\frac{105\pi a\sin\psi\cos\th}{16b}\biggr)\right] \,,
        \\ \CK_2^\bot&=-\frac{4Gm\w}{b} C_\b\biggr(\frac{L}{b}\biggr)^2\left[1+\frac{3a\sin\psi\sin\th}{b}+\frac{Gm}{b}\biggr(\frac{45\pi}{32}+\frac{15\pi a\sin\psi\sin\th}{2b}\biggr)\right] \,.
    \end{aligned} \label{eq:pho_wk_orth}
\end{align}
Note that beyond-general-relativity corrections are mostly $b^{-1}$ suppressed when compared with quantum corrections to the wavenumber kick ($\propto b^{-3}$)~\cite{Bjerrum-Bohr:2014zsa,Bai:2016ivl,Chi:2019owc}: although the leading $\b$ corrections scale as $\propto b^{-3}$, $\a'^2$ corrections scale as $\propto b^{-5}$. We may argue that this is an exception to the claims of ref.~\cite{Donoghue:1994dn} because electromagnetism is involved in the $FFR$ coupling.

We make some interesting observations. 
Firstly, the orthogonal wavenumber kick $\CK_2^\perp$ is non-vanishing even when $a=0$, which is generated from holomorphic impact parameters $\fb$ and $\bfb$ in the Magnusian matrices \eqref{eq: Photon Eik G1} and \eqref{eq: Photon Eik G2}. 
Secondly, the wavenumber kick matrix $\CK^i$ is diagonalised when we set $C_\b = 0$ ($\b = 0$), which implies that photons acquire different wavenumber kicks depending on their helicity: the incoming ray of light is split into two distinct rays, similar to how a beam of electrons in the Stern--Gerlach experiment splits into two distinct beams depending on the spin quantum numbers. 

The behaviour of the wavenumber kick for the case $C_\b \neq 0$ is qualitatively different: the independent components of the wavenumber kick do not commute,
\begin{align}
    [\CK^\para,\CK^\bot] \neq 0 \;,
\end{align}
which means the two matrices $\CK^\para$ and $\CK^\bot$ \emph{cannot} be simultaneously diagonalised and the incoming ray of light will disperse into a blob of rays. 
Note that this effect is still present for the non-spinning case $a=0$, which has been overlooked in previous studies~\cite{AccettulliHuber:2020oou}. The neglection can be traced to the order of operations: we compute the wavenumber kick first and then attempt to diagonalise the wavenumber kick (which is not possible), while in ref.~\cite{AccettulliHuber:2020oou} the Magnusian matrix was diagonalised first (which is possible for $a = 0$) and then the wavenumber kick is computed as its derivatives. The two operations do not commute because the similarity transform that can simultaneously diagonalise the matrices \eqref{eq: Photon Eik G1} and \eqref{eq: Photon Eik G2} requires the transformation matrix to depend on the impact parameter.

\begin{figure}[h]
    \centering
    \includegraphics[width=0.6\textwidth]{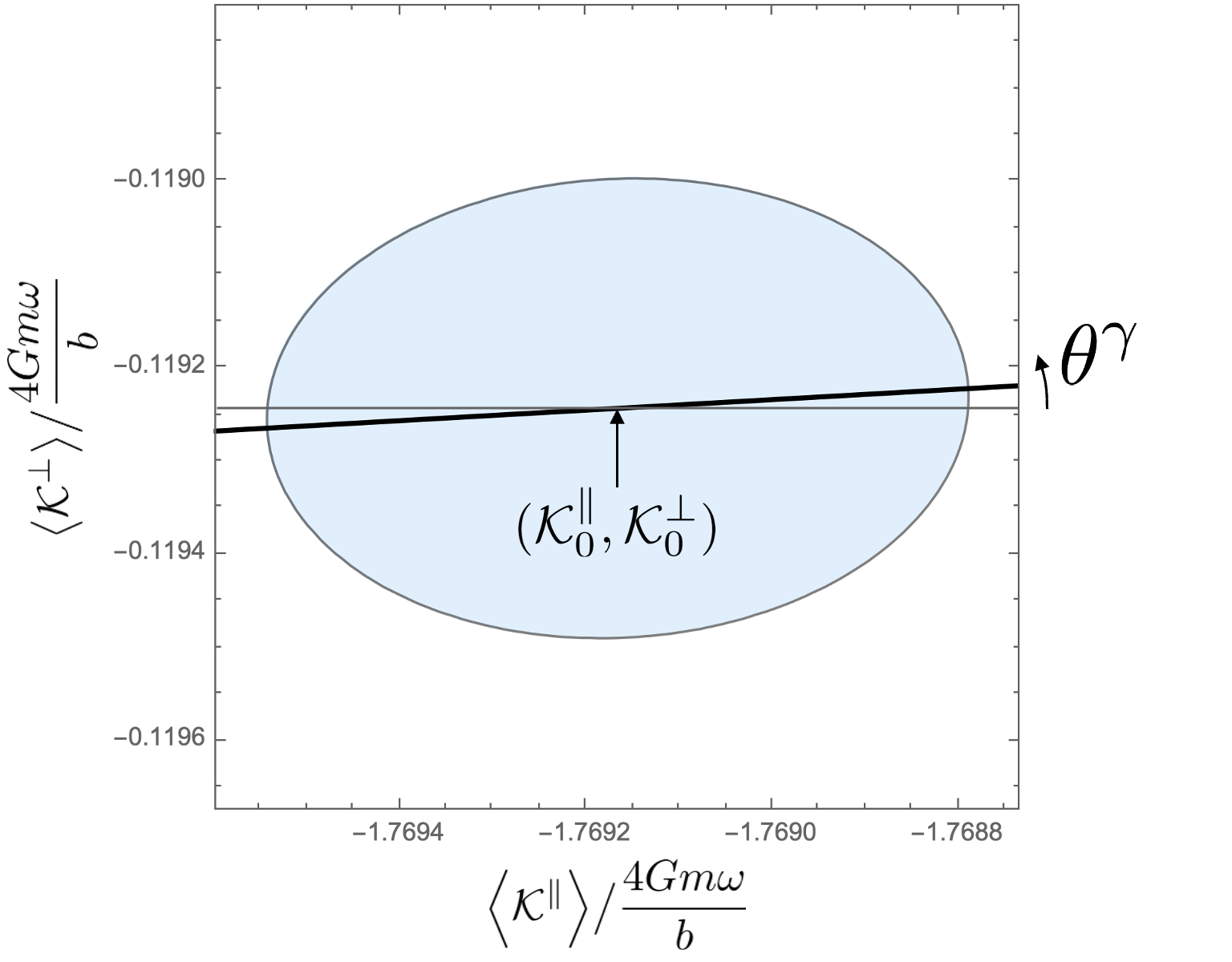}
    \caption{Expectation values of the wave number kick matrices \eqref{eq:pho_wk_para} and \eqref{eq:pho_wk_orth}
    for all possible helicity configurations $\Psi$ appear as an elliptic region in the wavenumber space. The centre of the ellipse sits at $(\CK_0^\para , \CK_0^\perp)$. The following parameter choices are used to plot the figure: $C_\b=C_{\a'^2}=1$, $L=0.05$, $\l=0.1$, $a=0.5$, $Gm=1$, $b=5$, $\theta=\pi/3$, and $\psi=3\pi/10$.
    }
    \label{fig:pho_wk}
\end{figure}
Fig.~\ref{fig:pho_wk} is the expectation value of the wavenumber kick plotted as a function of polarisation specified by the state $\Ket{\Psi}$, which is constructed from the positive (negative) helicity state $\Ket{+}$ ($\Ket{-}$) by the linear combination
\begin{align}
    \Ket{\Psi} = \cos \frac{\Omega}{2} \Ket{+} + e^{i \Phi} \sin \frac{\Omega}{2} \Ket{-} \,. \label{eq:polarisation_selection}
\end{align}
The parameters lie in the range $\Omega \in [ 0 , \pi]$ and $\Phi \in [0 , 2 \pi)$. 
If the two matrices $\CK^\para$ and $\CK^\bot$ can be diagonalised simultaneously, the expectation values will form a line where the endpoints correspond to the simultaneous eigenvalues. On the contrary, the expectation values plotted in fig.~\ref{fig:pho_wk} form a solid ellipse. We also observe that the semi-major axis of the ellipse forms an angle
\begin{align}
    \th^{\g}= \frac{a}{b} \left(\frac{256\cos\psi}{45\pi}+\frac{5\sin\psi\cos\th}{6}+\frac{Gm}{b}\left(6\cos\psi+\frac{225\pi\sin\psi\cos\th}{256}\right)\right)
\end{align}
with respect to the impact parameter due to spin effects. 
The ellipse in fig.~\ref{fig:pho_wk} collapses to a line when $C_\b \to 0$, consistent with the fact that the matrices $\CK^\para$ and $\CK^\bot$ are diagonalised in this limit.

\subsection{Graviton scattering} \label{sec:gravsc}
The tree-level Compton amplitude is computed as a sum of two contributions,
\begin{align} \label{eq: Graviton tree amp}
\begin{aligned}
    &\left. M_4^{h} \right|_{G^1}\sim
    \quad
\begin{fmffile}{gravtion_LO_EH}
\parbox{80pt}{
    \begin{fmfgraph*}(80,60)
        \fmfstraight
        \fmfleft{i1,c1,o1}
        \fmfright{i2,c2,o2}
        \fmf{phantom}{c1,b1,e1,e2,e3,b2,c2}
         \fmffreeze
        \fmf{plain,width=2}{i1,b1,o1}
        \fmf{dbl_wiggly}{b1,b2}
        \fmf{dbl_wiggly}{i2,b2,o2}
        \fmfv{decor.shape=circle,decor.filled=shaded,decor.size=30}{b1}
        \fmfv{decor.shape=circle,decor.filled=empty,decor.size=30,label=EH,label.dist=0}{b2}
        \fmf{phantom}{i1,d1,i2}
        \fmf{phantom}{o1,d2,o2}
        \fmf{dashes,tension=2,rubout=5}{d1,d2}
    \end{fmfgraph*}
}
\end{fmffile}
    \quad + \quad 
\begin{fmffile}{graviton_LO_R3}
\parbox{80pt}{
    \begin{fmfgraph*}(80,60)
        \fmfstraight
        \fmfleft{i1,c1,o1}
        \fmfright{i2,c2,o2}
        \fmf{phantom}{c1,b1,e1,e2,e3,b2,c2}
         \fmffreeze
        \fmf{plain,width=2}{i1,b1,o1}
        \fmf{dbl_wiggly}{b1,b2}
        \fmf{dbl_wiggly}{i2,b2,o2}
        \fmfv{decor.shape=circle,decor.filled=shaded,decor.size=30}{b1}
        \fmfv{decor.shape=circle,decor.filled=empty,decor.size=30,label=$R^3$,label.dist=0}{b2}
        \fmf{phantom}{i1,d1,i2}
        \fmf{phantom}{o1,d2,o2}
        \fmf{dashes,tension=2,rubout=5}{d1,d2}
    \end{fmfgraph*}
}
\end{fmffile}
    \\ & = \left(\frac{\k}{2}\right)^2 \frac{4m^2\w^2}{|\vec{q}|^2}\left[\left(1-\frac{i(n\cdot S)}{m^2\w}\right)\left(I_{2\times 2}+\frac{\a'^2}{16}\begin{pmatrix}
        0&\fq^4\\\bfq^4&0
    \end{pmatrix}\right)+\frac{1}{16\w^4}\left(1+\frac{q\cdot S}{m}\s_3\right)\begin{pmatrix}
        0&\fq^4\\\bfq^4 &0
    \end{pmatrix}\right] \;,
\end{aligned}
\end{align}
where matrix product is implied for the $\s_3$ appearing at the end of the equation.
The corresponding Magnusian matrix is\footnote{The relations $\fd=i\fa$ and $\bfd=-i\bfa$ satisfied by the holomorphic variables were used to remove $i$.}
\begin{align} \label{eq: Graviton Eik G1}
    \begin{split}
        \chi_{G^1}^{h} = 4 m\w G&\left[\left(\log \frac{b_0}{b}-\frac{\vec{d}\cdot\vec{b}}{b^2}\right)I_{2\times 2}+\frac{3}{16\w^4\bfb^4}\left(1+\frac{4\bfd}{\bfb}\right)\s_++\frac{3}{16\w^4\fb^4}\left(1+\frac{4\fd}{\fb}\right)\s_-\right.
        \\&\quad\left.+ \a'^2 \left( \frac{3}{16\bfb^4}\left(1-\frac{4\bfd}{\bfb}\right)\s_+ + \frac{3}{16\fb^4}\left(1-\frac{4\fd}{\fb}\right)\s_- \right)\right]\;.
    \end{split}
\end{align}
Unlike the photon case \eqref{eq: Photon Eik G1}, helicity-flipping amplitudes are non-vanishing for graviton scattering. However, their effects are highly suppressed ($\propto \e_{\text{wo}}^4 = (\w b)^{-4}$) as wave-optics corrections and we will neglect them in observable computations. 

The one-loop Compton amplitude is computed as a sum of three contributions,
\begin{align}
\begin{aligned}
    \left. M^{h}_{4} \right|_{G^2}\sim&\quad
    \begin{fmffile}{graviton_NLO_EH}
    \parbox{80pt}{
        \begin{fmfgraph*}(80,60)
            \fmfstraight
            \fmfleft{i1,c1,o1}
            \fmfright{i2,c2,o2}
            \fmf{phantom}{c1,b1,e1,e2,e3,b2,c2}
            \fmf{phantom}{i1,d1,i2}
            \fmf{phantom}{o1,d2,o2}
            \fmf{phantom}{d1,f1,f2,f3,d2}
             \fmffreeze
            \fmf{plain,width=2}{i1,b1,o1}
            \fmf{dbl_wiggly,left=0.7}{b1,b2}
            \fmf{dbl_wiggly,right=0.7}{b1,b2}
            \fmf{dbl_wiggly}{i2,b2,o2}
            \fmfv{decor.shape=circle,decor.filled=shaded,decor.size=30}{b1}
            \fmfv{decor.shape=circle,decor.filled=empty,decor.size=30,label=EH,label.dist=0}{b2}
            \fmf{dashes,rubout=5}{d1,f1}
            \fmf{dashes,rubout=5}{d2,f3}
        \end{fmfgraph*}
    }
    \end{fmffile}
        \quad + \quad
    \begin{fmffile}{graviton_NLO_R3}
    \parbox{80pt}{
        \begin{fmfgraph*}(80,60)
            \fmfstraight
            \fmfleft{i1,c1,o1}
            \fmfright{i2,c2,o2}
            \fmf{phantom}{c1,b1,e1,e2,e3,b2,c2}
            \fmf{phantom}{i1,d1,i2}
            \fmf{phantom}{o1,d2,o2}
            \fmf{phantom}{d1,f1,f2,f3,d2}
             \fmffreeze
            \fmf{plain,width=2}{i1,b1,o1}
            \fmf{dbl_wiggly,left=0.7}{b1,b2}
            \fmf{dbl_wiggly,right=0.7}{b1,b2}
            \fmf{dbl_wiggly}{i2,b2,o2}
            \fmfv{decor.shape=circle,decor.filled=shaded,decor.size=30}{b1}
            \fmfv{decor.shape=circle,decor.filled=empty,decor.size=30,label=$R^3$,label.dist=0}{b2}
            \fmf{dashes,rubout=5}{d1,f1}
            \fmf{dashes,rubout=5}{d2,f3}
        \end{fmfgraph*}
    }
    \end{fmffile}\quad+\quad \begin{fmffile}{graviton_NLO_R4}
    \parbox{80pt}{
        \begin{fmfgraph*}(80,60)
            \fmfstraight
            \fmfleft{i1,c1,o1}
            \fmfright{i2,c2,o2}
            \fmf{phantom}{c1,b1,e1,e2,e3,b2,c2}
            \fmf{phantom}{i1,d1,i2}
            \fmf{phantom}{o1,d2,o2}
            \fmf{phantom}{d1,f1,f2,f3,d2}
             \fmffreeze
            \fmf{plain,width=2}{i1,b1,o1}
            \fmf{dbl_wiggly,left=0.7}{b1,b2}
            \fmf{dbl_wiggly,right=0.7}{b1,b2}
            \fmf{dbl_wiggly}{i2,b2,o2}
            \fmfv{decor.shape=circle,decor.filled=shaded,decor.size=30}{b1}
            \fmfv{decor.shape=circle,decor.filled=empty,decor.size=30,label=$R^4$,label.dist=0}{b2}
            \fmf{dashes,rubout=5}{d1,f1}
            \fmf{dashes,rubout=5}{d2,f3}
        \end{fmfgraph*}
    }
    \end{fmffile}\quad.
\end{aligned}
\end{align}
Similar to the photon case \eqref{eq: Photon Eik G2}, only the triangle contribution contributes to the eikonal phase / Magnusian because exponentiation of the tree-level contribution reproduces the box and the crossed-box contributions. The Magnusian matrix is
\begin{align} \label{eq: Graviton Eik G2}
\begin{split}
    \chi_{G^2}^{h} &=\frac{15\pi G^2m^2\w}{4b}\left[\left(1-\frac{4(\vec{d}\cdot\vec{b})}{3b^2}\right)I_{2\times 2}-\frac{2(\hat{k}_2\cdot\vec{a})}{3\w b^2}\s_3\right.
    \\&\phantom{SPA}+\a'^2 \left\{ -\left(\frac{3}{80b^4}+\frac{1}{\w^2 b^2}\frac{105}{64b^4}\right)I_{2\times 2}\right.
    \\&\phantom{SPACESP}+\left(\frac{91}{256\bfb^4}-\frac{7(81\fb\bfd+11\bfb\fd)}{64b^2\bfb^4}+\frac{1}{\w^2b^2}\left(\frac{189}{1024\bfb^4}+\frac{63(77\fb\bfd+27\bfb\fd)}{256b^2\bfb^4}\right)\right)\s_+
    \\&\phantom{SPACESP}+\left.\left(\frac{91}{256\fb^4}-\frac{7(81\bfb\fd+11\fb\bfd)}{64b^2\fb^4}+\frac{1}{\w^2b^2}\left(\frac{189}{1024\fb^4}+\frac{63(77\bfb\fd+27\fb\bfd)}{256b^2\fb^4}\right)\right)\s_- \right\}
    \\&\phantom{SP}+\tilde{\z}\;\;\;\left\{\left(\frac{21\w^2}{4b^4}-\frac{21\w^2(\vec{d}\cdot\vec{b})}{b^6}\right)I_{2\times 2}+\frac{105\w(\hat{k}_2\cdot\vec{a})}{2b^6}\s_3\right\}
    \\&\phantom{SP}+\z^-\left(\frac{21\w^2}{64\bfb^4}-\frac{21\w^2(18\fb\bfd-2\bfb\fd)}{64b^2\bfb^4}+\frac{1}{\w^2b^2}\left(\frac{945\w^2}{128\bfb^4}-\frac{945\w^2(11\fb\bfd-3\bfb\fd)}{64b^2\bfb^4}\right)\right)\s_+
    \\&\phantom{SP}+\z^+\left.\left(\frac{21\w^2}{64\fb^4}-\frac{21\w^2(18\bfb\fd-2\fb\bfd)}{64b^2\fb^4}+\frac{1}{\w^2b^2}\left(\frac{945\w^2}{128\fb^4}-\frac{945\w^2(11\bfb\fd-3\fb\bfd)}{64b^2\fb^4}\right)\right)\s_-\right]\;.
\end{split}
\end{align}
where we have included leading wave-optics corrections ($\e_{\text{wo}} = (\w b)^{-1}$) to all interactions. In the following sections, we assume the absence of the parity-odd interaction $\CC\tilde{\CC}$, so that $\z^+=\z^-=\z$, and compute the observables.

\subsubsection{Gravitational Faraday rotation}
The polarisation rotation angle \eqref{Pol rotation angle} arises at one-loop order \eqref{eq: Graviton Eik G2},
\begin{align} \label{eq:gravitonGFR}
\begin{aligned}
    \a_{h} &= \frac{5\pi G^2m^2  (\hat{k}_2 \cdot \vec{a})}{4b^3}\left(1-\frac{315(b\w)^2C_{\tilde{\z}}}{4}\left(\frac{L}{b}\right)^6\right)
    \\ &=\frac{5\pi G^2m^2a\cos\psi}{4b^3}\left(1-\frac{315}{4} \left( \frac{L}{\l} \right)^2 \left(\frac{L}{b}\right)^4 C_{\tilde{\z}} \right) \;,    
\end{aligned}
\end{align}
where it is found that the parity-even part of the $R^4$ interactions modify the rotation angle. This is qualitatively different from the photon case \eqref{eq:photonGFR} where EFT corrections to general relativity does not affect the rotation angle. Moreover, the rotation angle \eqref{eq:gravitonGFR} depends on the energy (or equivalently, the wavelength) of the graviton, implying that finite-size effects related to the wavepacket size affects the frame-dragging rate experienced by the graviton. We remark that the $R^4$ correction is also highly suppressed in $b^{-1}$ counting when compared to quantum corrections from general relativity ($\propto b^{-4}$)~\cite{Kim:2022iub}, confirming the validity of using quantised general relativity to make universal quantum gravity predictions~\cite{Donoghue:1994dn}. 

\subsubsection{Gravitational spin Hall effect}
The parallel components of the wavenumber kick are 
\begin{align}
    \begin{split}
        \CK_0^{\para}&=-\frac{4Gm\w}{b}\left[1+\frac{a\sin\psi\sin\th}{b}+\frac{Gm}{b}\left(\frac{15\pi}{16}+\frac{5\pi a\sin\psi\sin\th}{2b}\right)\right.
        \\&\left.\phantom{SPACESP}-C_{\a'^2}\biggr(\frac{L}{b}\biggr)^4\frac{45\pi Gm}{256b}\right.
        \\&\left.\left.\phantom{SPACESP}+C_{\tilde{\z}}\left(\frac{\w^2L^6}{b^4}\right)\biggr(\frac{Gm}{b}\biggr)\left(\frac{1575\pi}{64}+\frac{945\pi a\sin\psi\sin\th}{8b}\right)\right)\right]\;,
        \\\CK^{\para}_1&=-\frac{4Gm\w}{b}\left[C_{\a'^2}\biggr(\frac{L}{b}\biggr)^4\left(12+\frac{60a\sin\psi\sin\th}{b}+\biggr(\frac{Gm}{b}\biggr)\left(\frac{6825\pi}{256}+\frac{7245\pi a\sin\psi\sin\th}{32b}\right)\right)\right.
        \\&\left.\left.\phantom{SPACESP}+C_\z\left(\frac{\w^2L^6}{b^4}\right)\biggr(\frac{Gm}{b}\biggr)\left(\frac{1575\pi}{64}-\frac{945\pi a\sin\psi\sin\th}{8b}\right)\right.\right]\;,
        \\\CK^{\para}_2&=-\frac{4Gm\w}{b}\left[C_{\a'^2}\biggr(\frac{L}{b}\biggr)^4\left(\frac{60a\sin\psi\cos\th}{b}+\biggr(\frac{Gm}{b}\biggr)\left(\frac{192a\cos\psi}{b}+\frac{11025\pi a\sin\psi\cos\th}{64b}\right)\right)\right.
        \\&\left.\phantom{SPACES}\;-C_\z\left(\frac{\w^2L^6}{b^4}\right)\biggr(\frac{Gm}{b}\biggr)\frac{4725\pi a\sin\psi\cos\th}{32b}\right]\;,
        \\\CK^{\para}_3&= \frac{4Gm\w}{b}\times\frac{Gm}{b}\times\frac{1}{\w b}\times\frac{30a\cos\psi}{16b}\,,
    \end{split} \label{eq:grav_wk_para}
\end{align}
and the orthogonal components are
\begin{align}
    \begin{split}
        \CK_0^{\perp}&=-\frac{4Gm\w}{b}\left[\frac{a\sin\psi\cos\th}{b}+\biggr(\frac{Gm}{b}
        \biggr)\left(\frac{4 a\cos\psi}{b}+\frac{5\pi a\sin\psi\cos\th}{4b}\right)\right.
        \\&\left.\phantom{SPACESP}+C_{\tilde{\z}}\left(\frac{\w^2L^6}{b^4}\right)\biggr(\frac{Gm}{b}\biggr)\frac{315\pi a\sin\psi\cos\th}{16b}\right]\;,
        \\\CK^\perp_1&=-\frac{4Gm\w}{b}\left[C_{\a'^2}\biggr(\frac{L}{b}\biggr)^4\left(\frac{60a\sin\psi\cos\th}{b}+\biggr(\frac{Gm}{b}\biggr)\left(\frac{288a\cos\psi}{b}+\frac{9765\pi a\sin\psi\cos\th}{64b}\right)\right)\right.
        \\&\left.\phantom{SPACESP}-C_{\z}\left(\frac{\w^2L^6}{b^4}\right)\biggr(\frac{Gm}{b}\biggr)\frac{945\pi a\sin\psi\cos\th}{8b}\right]\;,
        \\\CK_2^\perp&=\frac{4Gm\w}{b}\left[C_{\a'^2}\biggr(\frac{L}{b}\biggr)^4\left(12+\frac{60a\sin\psi\sin\th}{b}+\biggr(\frac{Gm}{b}\biggr)\left(\frac{1365\pi}{64}+\frac{22995\pi a\sin\psi\sin\th}{128b}\right)\right)\right.
        \\&\left.\phantom{SPACESP}+C_{\z}\left(\frac{\w^2L^6}{b^4}\right)\biggr(\frac{Gm}{b}\biggr)\left(\frac{315\pi}{16}-\frac{6615\pi a\sin\psi\sin\th}{64b}\right)\right]\;.
    \end{split} \label{eq:grav_wk_orth}
\end{align}
In contrast to the photon case, all beyond-general-relativity effects are subleading in the $b^{-1}$ expansion when compared to the quantum corrections to the wavenumber kick ($\propto b^{-3}$)~\cite{Bai:2016ivl,Chi:2019owc}: the effects related to the UV completion of general relativity are suppressed compared to quantum effects from graviton exchange over long distances, therefore quantised general relativity can be used to make universal quantum gravity predictions~\cite{Donoghue:1994dn}.

\begin{figure}[h]
    \centering
    \includegraphics[width=0.6\textwidth]{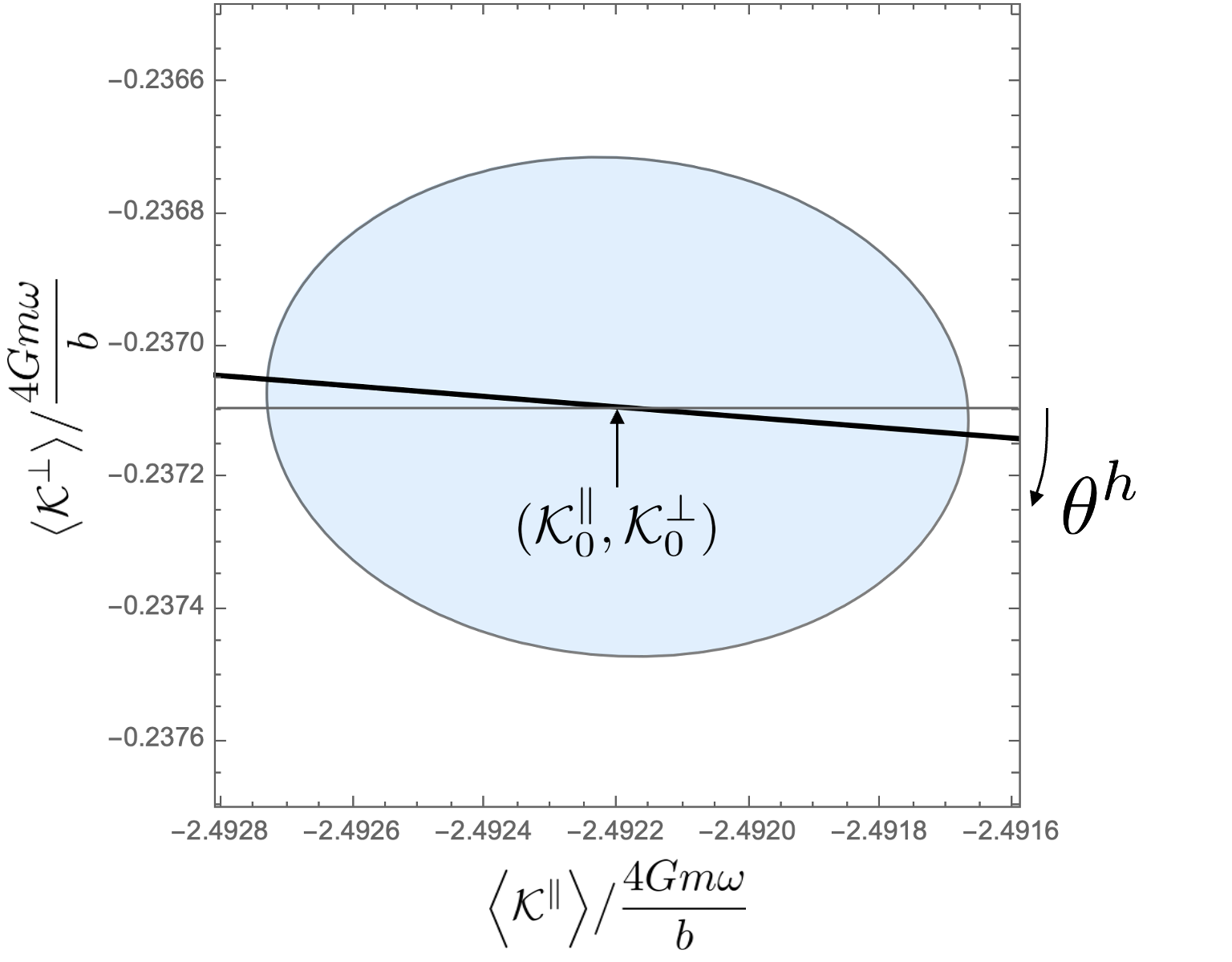}
    \caption{Expectation values of the wave number kick matrices \eqref{eq:grav_wk_para} and \eqref{eq:grav_wk_orth} for all possible helicity configurations $\Psi$ appear as an elliptic region in the wavenumber space. The centre of the ellipse sits at $(\CK_0^\para , \CK_0^\perp)$. The following parameter choices are used to plot the figure: $C_{\a'^2}=C_{\tilde{\z}}=C_{\z}=1$, $L=0.1$, $\l=0.02$, $a=0.5$, $Gm=1$, $b=3$, $\theta=\pi/3$, and $\psi=3\pi/7$. Contrary to fig.~\ref{fig:grav_wk2}, the parameters violating the hierarchy \eqref{Hierarchy of parameters} were chosen to make the elliptical shape prominent. The difference in the ``tilt'' from fig.~\ref{fig:pho_wk} is due to a different choice of $\psi$.
    }
    \label{fig:grav_wk}
\end{figure}
\begin{figure}[h]
    \centering
    \includegraphics[width=0.6\textwidth]{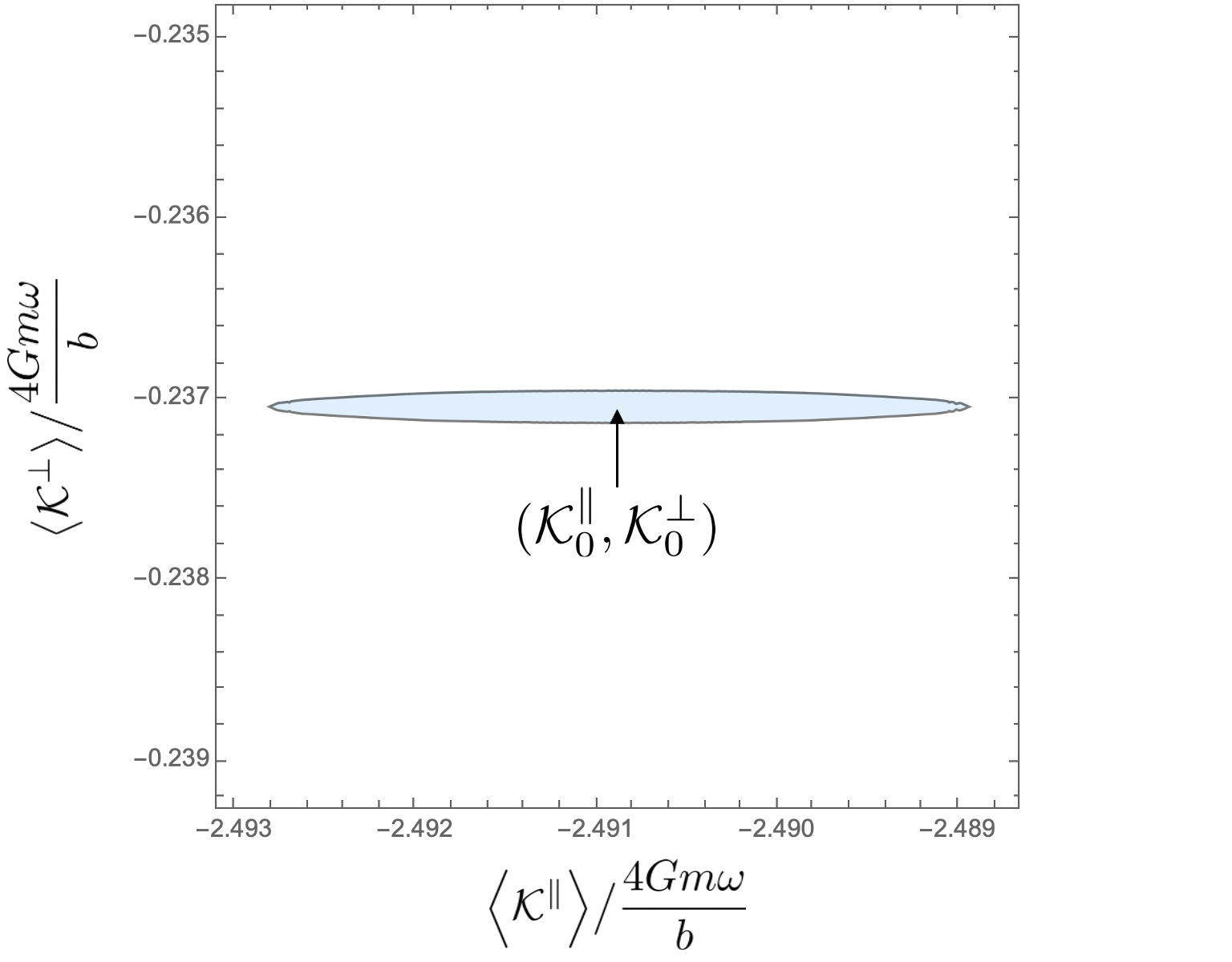}
    \caption{The same plot as fig.~\ref{fig:grav_wk} but with parameters obeying the hierarchy \eqref{Hierarchy of parameters}: $C_{\a'^2}=C_{\tilde{\z}}=C_{\z}=1$, $L=0.1$, $\l=0.25$, $a=0.5$, $Gm=1$, $b=3$, $\theta=\pi/3$, and $\psi=3\pi/7$. Contrary to fig.~\ref{fig:grav_wk}, the wave-optics correction $\CK_3^\para$ suppresses the characteristic elliptical shape arising from EFT corrections.
    }
    \label{fig:grav_wk2}
\end{figure}

Similarly to the photon case, the wavenumber kick matrices $\CK^i$ are diagonalised for general relativity ($C_{\a'^2}=C_{\z}=0$ or $\a'^2=\z=0$) and the independent components do not commute. 
Parametrising the graviton polarisation as \eqref{eq:polarisation_selection}, we can plot the expectation value of the wavenumber kick as in figs.~\ref{fig:grav_wk}-\ref{fig:grav_wk2}. 
One qualitative difference from the photon case (fig.~\ref{fig:pho_wk}) is the higher suppression of the EFT corrections ($\e_{\text{EFT}}^{4} = \frac{L^4}{b^4}$ compared to $\e_{\text{EFT}}^{2} = \frac{L^2}{b^2}$ of the photon case), which makes the wave-optics correction $\CK_3^\para$ the dominant feature of the plot. We chose to violate the hierarchy \eqref{Hierarchy of parameters} in fig.~\ref{fig:grav_wk} so that the effects of ``noncommutative'' wavenumber kick from EFT corrections are clearly visible, in contrast to fig.~\ref{fig:grav_wk2} where the EFT corrections are overshadowed by wave-optics corrections. 

Interestingly, the ``tilt'' of the ellipse is determined by the $R^3$ coupling and is independent of $R^4$ corrections: 
\begin{align}
    \;\theta^{h}=\begin{cases}
        \frac{a}{b}\left(\frac{4096\cos\psi}{455\pi}+\frac{17\sin\psi\cos\th}{26}+\frac{Gm}{b}\left(10\cos\psi+\frac{7245\pi\sin\psi\cos\th}{4096}\right)\right)&(\a'^2\neq 0,\;\z=0)
        \\\CO((a/b)^2)&(\a'^2=0,\;\z\neq 0)
    \end{cases}
\end{align}

\section{Time delay and spin-refined IR causality analysis} \label{sec:TD_causality}
It is found that the subleading bracket term of \eqref{def: Time delay} vanishes to the approximation order we are considering, 
\begin{align}
    \frac{1}{2!} \bigr\{ \chi_{G^1}, \{ \chi_{G^1},X_2^0 \} \bigr\} = \CO ( C_\#^2 ) \,,
\end{align}
for both photon and graviton scattering. 
Therefore, it suffices to diagonalise the energy derivative of the Magnusian matrix,
\begin{align}
    \D t={\rm Spec}\biggr\{\frac{\p}{\p\w}\bigr(\chi_{G^1}+\chi_{G^2}\bigr)\biggr\} \;,
\end{align}
when performing causality analysis. We will neglect wave-optics corrections ($\e_{\text{wo}} = (\w b)^{-1}$) in the analyses.\footnote{Note that the leading wave-optics corrections in general relativity do not contribute to the time delay because the corrections scale as $\propto \w^0$. This is consistent with the fact that gravitational Faraday rotation---the leading wave-optics correction contributes to gravitational Faraday rotation---is a geometric quantity independent of the massless particle's energy.}

In general, violations of causality constraints should be understood as insufficiency of using the action \eqref{The EFT} to describe the system: further UV information such as new degrees of freedom or higher dimensional operators should be included in the action to describe the system faithfully~\cite{Camanho:2014apa}.

\subsection{Photon scattering} 
The time delay for photon scattering is evaluated from \eqref{eq: Photon Eik G1} and \eqref{eq: Photon Eik G2} as
\begin{align}
    \begin{split}
        \D t^\g_{\pm} 
        =\;&4Gm\left[-\log\frac{b}{b_0}+\frac{a\sin\psi\sin\th}{b}+\frac{Gm}{b}\left(\frac{15\pi}{16}+\frac{5\pi a \sin\psi\sin\th}{4b}
        \right)\right.
        \\&\phantom{SP}\;\;-C_{\a'^2}\left(\frac{L}{b}\right)^4\frac{9\pi Gm}{256b}
        \\&\phantom{SP}\;\;\pm\frac{C_\b}{2}\left(\frac{L}{b}\right)^2\left.\left(1+\frac{2a\sin\psi\sin\th}{b}+\frac{Gm}{b}\left(\frac{45\pi}{32}+\frac{45\pi a\sin\psi\sin\th}{8b}\right)\right)\right] - \t_0^\pm \,,
    \end{split} \label{eq:pho_td}
\end{align}
which are the two eigenvalues of the time delay matrix $\D t = \frac{\partial}{\partial \w} \overleftrightarrow{\chi}$. 
The constants $\t_0^\pm$ are related to the IR regulator $b_0$ and are fixed by the conditions $\D t^{\g}_\pm(b=b_0)=0$. 
The time delay \eqref{eq:pho_td} is consistent with eq.(4.75) of ref.~\cite{AccettulliHuber:2020oou}.
The IR causality condition \eqref{eq: IR causality} becomes
\begin{align}
    \D t^{\g}_\pm -
    \D t^{\g}_{\rm GR} 
    =4Gm\left[-\frac{Gm}{b}\frac{9\pi C_{\a'^2}}{256}\biggr(\frac{L}{b}\biggr)^4\pm\frac{C_\b \CZ^\b}{2}\biggr(\frac{L}{b}\biggr)^2\right]\gtrsim-\frac{1}{\w} \,, \label{eq:pho_IRcaus}
\end{align}
where $\CZ^\b$ is a tunable parameter that can be tuned by changing the rotation axis (changing the Euler angles $\th$ and $\psi$) of the black hole,
\begin{align}
    \CZ^\b := 1+\frac{2a\sin\psi\sin\th}{b}+\frac{Gm}{b}\left(\frac{45\pi}{32}+\frac{45\pi a\sin\psi\sin\th}{8b}\right) \;. \label{eq:zb_def}
\end{align}
The inequality \eqref{eq:pho_IRcaus} should be satisfied by both $\D t^\g_\pm$ to ensure that there is no resolvable time advance. 

Assuming $C_\# \sim \CO(1)$, the IR causality condition \eqref{eq:pho_IRcaus} is dominated by $C_\b$ due to the extra factor $\frac{Gm}{b}\left(\frac{L}{b}\right)^2 = \e_{\text{PM}} \e_{\text{EFT}}^2 \ll 1$ attached to $C_{\a'^2}$; consult \eqref{eq:eps_hier} for the classical expansion parameters of the setup. 
As a first analysis, we set $C_{\a'^2} = 0$ and interpret the inequality as the condition that must be satisfied by the coupling $\b = C_\b L^2$, which leads to the lowest upper bound $\b_{\text{max}}$ on its absolute value, 
\begin{align}
    |\b| = |C_\b| L^2 \le \b_\text{max} = \frac{L Gm}{2 \max \left[ \e_{\text{PM}}^2 \CZ^\b \right]} \le \frac{1}{2} \frac{1}{\w} \frac{b^2}{Gm \CZ^\b} \,, \label{eq:pho_bound_0}
\end{align}
where the lowest upper bound is obtained from the maximally energetic photons below the EFT cutoff $\w=\w_{\rm max}=\L$. The absolute value $|C_\b|$ appears in \eqref{eq:pho_bound_0} because one of the eigenvalues trivially satisfies the inequality \eqref{eq:pho_IRcaus} for $C_{\a'^2} = 0$.
When we use the coupling $\b$ to define the EFT cutoff (i.e. $|C_\b| = 1$), the inequality \eqref{eq:pho_bound_0} can be rewritten as 
\begin{align}
    |\b| = L^2 \le \frac{1}{4} \left( \frac{Gm}{\max \left[ \e_{\text{PM}}^2 \CZ^\b \right]} \right)^2 \,. \label{eq:pho_bound_1}
\end{align}
The spin-orientation dependence of $\max \left[ \e_{\text{PM}}^2 \CZ^\b \right]$ can be used to lower the upper bound \eqref{eq:pho_bound_1}, leading to the \emph{spin refinement} of the causality constraint.\footnote{\label{fn:PMcorr}The allowed values of the impact parameter $b \ge b_c$ depends on the configuration; see footnote \ref{fn:crit_b}. It is possible that the subleading post-Minkowskian corrections to the Magnusian/eikonal invalidate the one-loop time delay \eqref{eq:pho_IRcaus} as $b \to b_c$ and significantly modifies the bounds \eqref{eq:pho_bound_0} and \eqref{eq:pho_bound_1}~\cite{Alexander:2025gdn}, but this should be viewed as a limitation from the difficulty of higher-loop calculations rather than as a fundamental limitation of the time delay computed from scattering amplitudes.} As reference values, we provide some values of $\max \left[ \e_{\text{PM}}^2 \CZ^\b \right]$ for Kerr geodesics~\cite{Iyer:2009wa},
\begin{align}
    \max \left[ \e_{\text{PM}}^2 \CZ^\b \right] &\simeq \left\{
    \begin{aligned}
        & 6.85 \times 10^{-2} && & a&=0 \,,& & b=b_c \\
        & 4.65\times 10^{-2} && & a&=Gm \,,& &\text{retrograde } (b=b_c) \\
        & 3.29\times10^{-3} && & a&=Gm \,,& &\text{prograde} \\
        & 7.80 \times 10^{-2} && & a&= 0.476 \, Gm \,,& &\text{prograde } (b=b_c) \\
    \end{aligned}
    \right. \label{eq:eps2zb_ref}
\end{align}
Critical impact parameters were used for the non-spinning case, the extremal Kerr retrograde orbit, and the optimal prograde orbit ($a = 0.476 \, Gm$), while the maximum of $\e_{\text{PM}}^2 \CZ^\b$ in the range $b \ge b_c$ was used for the extremal Kerr ($a = Gm$) prograde orbit.\footnote{$\CZ^\b$ becomes negative for sufficiently small values of $b$ and invalidates the computation in this regime.} 
This gives an estimate of $\sim 30 \%$ enhancement in the causality constraint \eqref{eq:pho_bound_1} from optimal spin refinement. 

We remark that for \emph{fixed} values of impact parameter, retrograde orbits experience longer Shapiro time delays compared to prograde in general relativity, a fact that seems to be related to retrograde orbits giving tighter bounds than prograde orbits for extremal Kerr. The optimal bound arises from prograde orbits because the trajectories can be put closer to the black hole: the effect of smaller impact parameters dominates over the effect of reduction in time delay for prograde orbits.

If we set $C_\b = 0$ in \eqref{eq:pho_IRcaus} instead, we only get a meaningful constraint for $C_{\a'^2} > 0$. A similar analysis shows that the upper bound $\a'^2_{\text{max}}$ on the coupling $\a'^2 = C_{\a'^2} L^4$ is
\begin{align}
    \a'^2 = C_{\a'^2} L^4 \le \a'^2_{\text{max}} = \frac{64}{9 \pi} \frac{L(Gm)^3}{\e_{\text{UV}}^5} \le \frac{64}{9 \pi} \frac{1}{\w} \frac{b^5}{(Gm)^2} \,, \label{eq:pho_bound_2}
\end{align}
where we used the critical impact parameter $b= b_c$ to maximise $\e_{\text{PM}} \le \e_{\text{UV}}$. Using the coupling $\a'$ to define the EFT cutoff (i.e. $C_{\a'^2} = 1$) recasts the inequality \eqref{eq:pho_bound_2} as
\begin{align}
    \a'^2 = L^4 \le \a'^2_{\text{max}} = \left( \frac{64}{9 \pi} \right)^{\frac{4}{3}} \frac{(Gm)^4}{(\e_{\text{UV}})^{\frac{20}{3}}} \,, \label{eq:pho_bound_3}
\end{align}
which also shows spin refinement by \emph{two orders} of magnitude (a factor of $(\frac{3\sqrt{3}}{2})^{20/3}\sim 580$ enhancement) due to orientation dependence of $\e_{\text{UV}}$, when we use Kerr geodesics as reference values~\cite{Iyer:2009wa}
\begin{align}
    \e_{\text{UV}} = \frac{Gm}{b_c} = \left\{
    \begin{aligned}
        & 1/{3 \sqrt{3}} && & a&=0 \\
        & 1/{7} && & a&=Gm \,,\quad \text{retrograde} \\
        & 1/{2} && & a&=Gm \,,\quad \text{prograde} \\
    \end{aligned}
    \right. \label{eq:epsUV_extKerr}
\end{align}
We see that smaller black holes are more sensitive to EFT corrections, which seems to reflect the fact that spacetime curvature is larger for smaller black holes: the inequalities \eqref{eq:pho_bound_1} and \eqref{eq:pho_bound_3} bound $|\b|$ and $\a'^2$ more strongly when the horizon scale of the black hole ($Gm$) is smaller.

Since EFT cutoffs are usually given in units of energy, we also present the inequalities \eqref{eq:pho_bound_0} and \eqref{eq:pho_bound_2} as constraints on $\L = L^{-1}$ when the Wilson coefficients $C_\#$ are fixed as $\CO(1)$ numbers, following the analysis of ref.~\cite{deRham:2021bll}:
\begin{align}
    \L &\ge \L_{FFR} = \frac{2 |C_\b| \max\left[\e_{\text{PM}}^2 \CZ^\b_{c}\right]}{Gm} \,, && & & C_{\a'^2} = 0 \,,
    \\ \L &\ge \L_{R^3}^\g = \left( \frac{9 \pi}{64} \right)^{\frac{1}{3}} \frac{ \left( \e_{\text{UV}}^5 \, C_{\a'^2} \right)^{ \frac{1}{3}} }{Gm} \,, && & & C_\b = 0\,, \,C_{\a'^2} > 0 \,.
\end{align}

\subsection{Graviton scattering}
The time delay for graviton scattering is computed from \eqref{eq: Graviton Eik G1} and \eqref{eq: Graviton Eik G2} as 
\begin{align}
    \begin{split}
        \D t^h_\pm
        &= 4Gm\left[-\log\frac{b}{b_0}+\frac{a\sin\psi\sin\th}{b}+\frac{Gm}{b}\left(\frac{15\pi}{16}+\frac{5\pi a\sin\psi\sin\th}{4b}\right)\right.
        \\&\phantom{SP}+C_{\a'^2}\left(\frac{L}{b}\right)^4\left(-\frac{9\pi Gm}{256b} \right.
        \\ &\phantom{SPASDFASDFA}\left.\pm \left(3+\frac{12a\sin\psi\sin\th}{b}+\frac{Gm}{b}\left(\frac{1365\pi}{256}+\frac{2415\pi a\sin\psi\sin\th}{64b}\right)\right)\right.
        \\&\phantom{SP}+C_{\tilde{\z}}\left(\frac{\w^2L^6}{b^4}\right)\left(\frac{Gm}{b}\right)\left(\frac{945\pi}{64}+\frac{945\pi a\sin\psi\sin\th}{16}\right)
        \\&\phantom{SP}\pm C_\z\left(\frac{\w^2L^6}{b^4}\right)\left(\frac{Gm}{b}\right)\left(\frac{945\pi}{64}\left.-\frac{945\pi a\sin\psi\sin\th}{16}\right)\right] - \t_0^\pm \,,
    \end{split} \label{eq:grav_td}
\end{align}
where the constants $\t_0^\pm$ are fixed by the conditions $\D t^{h}_\pm(b=b_0)=0$, similarly to the photon case \eqref{eq:pho_td}. 
The computed time delay \eqref{eq:grav_td} is consistent with eq.(4.41) and eq.(4.48) of ref.~\cite{AccettulliHuber:2020oou}. 

Following the analysis for photon scattering, we will consider cases $C_{\a'^2} = 0$ and $C_{\z} = C_{\tilde{\z}} = 0$ separately. For $C_{\z} = C_{\tilde{\z}} = 0$, the IR causality condition \eqref{eq: IR causality} becomes
\begin{align}
    \D t^{h}_\pm - \D t^{h}_{\rm GR} = \mp 4GmC_{\a'^2}\biggr(\frac{L}{b}\biggr)^4 \CZ_\pm^{\a'^2} \gtrsim-\frac{1}{\w} \,, \label{eq:grav_IRcaus_1}
\end{align}
where $\CZ_\pm^{\a'^2}$ are tunable parameters defined as
\begin{align}
    \CZ_+^{\a'^2} &= 3\left(1+\frac{4a\sin\psi\sin\th}{b}\right)+\frac{Gm}{b}\left(\frac{687\pi}{128}+\frac{2415\pi a\sin\psi\sin\th}{64b}\right) \;,
    \\ \CZ_-^{\a'^2} &= 3\left(1+\frac{4a\sin\psi\sin\th}{b}\right)+\frac{Gm}{b}\left(\frac{339\pi}{64}+\frac{2415\pi a\sin\psi\sin\th}{64b}\right) \;.
\end{align}
Contrary to the bound from photon scattering \eqref{eq:pho_bound_2}, the inequality \eqref{eq:grav_IRcaus_1} imposes two-sided bounds on the coupling $\a'^2$;
\begin{align}
\begin{aligned}
    -\frac{1}{4}\frac{1}{\w}\frac{b^4}{Gm\CZ^{\a'^2}_-} & \le \a'^2_{\text{min}} = - \frac{L(Gm)^3}{4 \max \left[\e_{\rm PM}^4\CZ^{\a'^2}_{-}\right]} \le \a'^2 = C_{\a'^2} L^4 \,, && & \a'^2 &\le 0 \,,
    \\ \a'^2 = C_{\a'^2} L^4 &\le \a'^2_{\text{max}} = \frac{L(Gm)^3}{4 \max \left[\e_{\rm PM}^4 \CZ^{\a'^2}_{+}\right]} \le \frac{1}{4}\frac{1}{\w}\frac{b^4}{Gm\CZ^{\a'^2}_+} \,, && & \a'^2 &\ge 0 \,.
\end{aligned} \label{eq:grav_bound_1}
\end{align}
The bounds are found by tuning to maximally energetic gravitons below the EFT cutoff $\w=\w_{\rm max}=\L$. 
When the coupling $\a'^2$ defines the EFT cutoff (i.e. $|C_{\a'^2}| = 1$), the bound \eqref{eq:grav_bound_1} can be reformulated as
\begin{align}
    \a'^2_{\text{min}} = - \frac{1}{2^\frac{8}{3}} \frac{(Gm)^4}{\left( \max \left[\e_{\rm PM}^4\CZ^{\a'^2}_{-}\right] \right)^{\frac{4}{3}}} \le \a'^2 = \text{sgn}(C_{\a'^2}) L^4 \le \a'^2_{\text{max}} = \frac{1}{2^\frac{8}{3}} \frac{(Gm)^4}{\left( \max \left[\e_{\rm PM}^4\CZ^{\a'^2}_{+}\right] \right)^{\frac{4}{3}}} \,. \label{eq:grav_bound_2}
\end{align}
As reference values, we provide some values of $\e_{\text{UV}}^4 \CZ^{\a'^2}_{\pm}$ for Kerr geodesics~\cite{Iyer:2009wa},\footnote{See the paragraph below \eqref{eq:eps2zb_ref} for how the values were computed.} 
\begin{align}
    \max \left[ \e_{\text{PM}}^4 \CZ^{\a'^2}_{\pm} \right] &\simeq \left\{
    \begin{aligned}
        & 8.57\times 10^{-3} && & a&=0 \,,\,\a'^2\ge0 & & b=b_c\\
         & 8.51\times 10^{-3} && & a&=0 \,,\,\a'^2\le0 & & b=b_c\\
        & 3.97\times 10^{-3} && & a&=Gm \,,\, \a'^2 \ge 0 \,, & & \text{retrograde } (b=b_c) \\
        & 3.96\times 10^{-3} && & a&=Gm \,,\, \a'^2 \le 0 \,, & &  \text{retrograde } (b=b_c) \\
        & 5.39\times 10^{-4} && & a&=Gm \,,\, \a'^2 \ge 0 \,, & & \text{prograde} \\
        & 5.23\times 10^{-4} && & a&=Gm \,,\, \a'^2 \le 0 \,, & &  \text{prograde} \\
        & 9.40 \times 10^{-3} && & a&=0.246 \, Gm \,,\, \a'^2 \ge 0 \,, & & \text{prograde } (b=b_c) \\
        & 9.30 \times 10^{-3} && & a&=0.242 \, Gm \,,\, \a'^2 \le 0 \,, & &  \text{prograde } (b=b_c) \\
    \end{aligned}
    \right.
\end{align}
which gives an estimate of $\sim 10 \%$ enhancement in the causality constraint \eqref{eq:grav_bound_2} from optimal spin refinement. 
Similarly to the photon case, the inequality \eqref{eq:grav_bound_1} can be rewritten as a constraint on $\L = L^{-1}$ when the Wilson coefficient $C_{\a'^2}$ is fixed as a $\CO(1)$ number~\cite{deRham:2021bll}:
\begin{align}
    \L \ge \L_{R^3,\pm}^h = \frac{\left( 4 |C_{\a'^2}| \max\left[\e^4_{\rm PM}\CZ^{\a'^2}_\pm\right]\right)^{\frac{1}{3}}}{Gm} \,, && & & C_{\z} = C_{\tilde{\z}} = 0 \,.
\end{align}

\begin{figure}[h]
    \centering
    \includegraphics[width=0.6\textwidth]{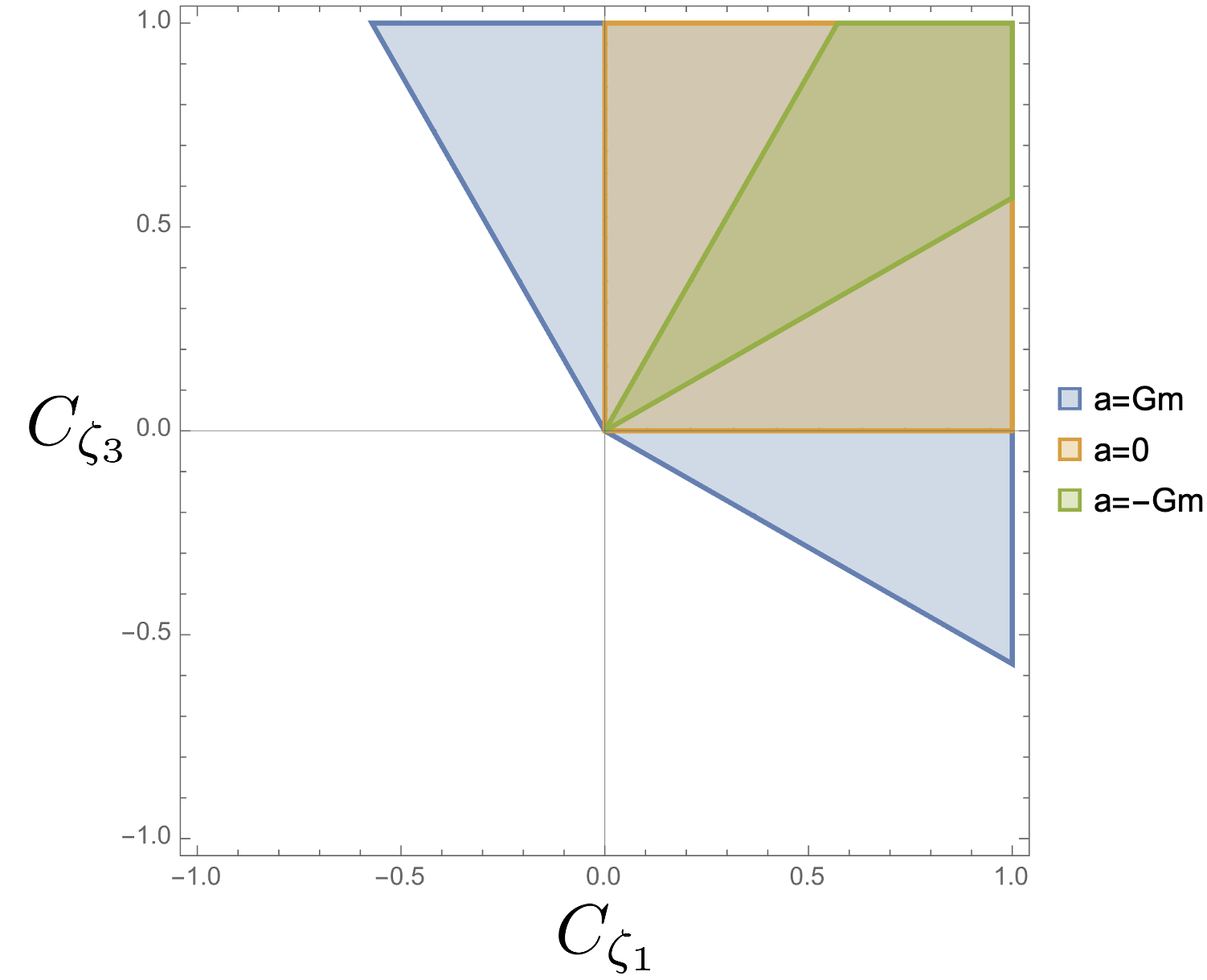}
    \caption{
    Allowed regions for $C_{\z_1} =\frac{1}{8}(C_{\tilde{\z}}+C_\z)$ and $C_{\z_3} =\frac{1}{8}(C_{\tilde{\z}}-C_\z)$ given by \eqref{eq:grav_IRcaus_3} with $b = 7Gm$. The prograde orbit ($a = - Gm$) narrows the allowed regions, while the retrograde orbit ($a = Gm$) expands the allowed regions. The spin effects favour correlated $C_{\z_1}$ and $C_{\z_3}$. When small negativity is allowed as in \eqref{eq:grav_IRcaus_2}, the negativity shifts the tip of the cone to the lower left along the diagonal line. In the regime of EFT validity $L \ll Gm$, the tip sits far outside the range of the plot $[-1,+1] \times [-1,+1]$ and no meaningful constraints exist for the $\CO(1)$ Wilson coefficients, unless a smaller value of $b \sim 4Gm$ is chosen so that the cone becomes sufficiently narrow.
    }
    \label{fig:R4caus}
\end{figure}

For the other case $C_{\a'^2}=0$, the IR causality condition \eqref{eq: IR causality} becomes
\begin{align}
    \D t^{h}_\pm - \D t^{h}_{\rm GR} = \frac{945 \pi}{16} Gm(b\w)^2\biggr(\frac{L}{b}\biggr)^6\frac{Gm}{b} \left[ \left( C_{\tilde{\z}} \pm C_{\z} \right) + \left( C_{\tilde{\z}} \mp C_{\z} \right) \frac{4a \sin\psi\sin\th}{b} \right]\gtrsim-\frac{1}{\w} \,. \label{eq:grav_IRcaus_2}
\end{align}
Since we have two parameters $C_{\tilde\z}$ and $C_{\z}$ to be bounded, we present 2d plots of the allowed regions. The small allowed negativity of \eqref{eq:grav_IRcaus_2} invalidates any meaningful constraints on $C_{\tilde\z} \sim \CO(1)$ and $C_{\z} \sim \CO(1)$ in the regime of EFT validity $L \ll Gm$, therefore we instead impose strict nonnegativity and consider the condition
\begin{align}
    \left( C_{\tilde{\z}} \pm C_{\z} \right) + \left( C_{\tilde{\z}} \mp C_{\z} \right) \frac{4a}{b} \ge 0 \,, \label{eq:grav_IRcaus_3}
\end{align}
for the plot in fig.~\ref{fig:R4caus}. Although the plot \emph{cannot} be interpreted as valid bounds on the Wilson coefficients, the shrinking of the allowed regions can be interpreted as evidence of refinement of causality bounds from considering spin effects.

Fig.~\ref{fig:R4caus} shows the allowed regions for the Wilson coefficients $C_{\z_1} =\frac{1}{8}(C_{\tilde{\z}}+C_\z)$ and $C_{\z_3} =\frac{1}{8}(C_{\tilde{\z}}-C_\z)$, for $b = 7Gm$ and different values of $a$. The choice $b = 7Gm$ is based on the critical impact parameter of extremal Kerr retrograde orbits \eqref{eq:epsUV_extKerr}. The tighter bounds come from prograde orbits ($a = - Gm$), which narrows the cone of allowed regions. Considering that prograde orbits allow smaller values of the impact parameter, it is possible that the cone can be shrunk even further by choosing $b < 7Gm$ (even a straight line for $b = 4Gm$), refining the independent conditions $C_{\z_1} \ge 0$ and $C_{\z_3} \ge 0$ for the non-spinning case to the correlated condition $C_{\z_1} \simeq C_{\z_3} \ge 0$ from spin effects. 

Note that this narrowing effect will also be present when we use the causality condition \eqref{eq:grav_IRcaus_2}: up to the caveat of higher-loop (post-Minkowskian) corrections invalidating the time delay \eqref{eq:grav_IRcaus_2} for small impact parameters,\footnote{See footnote~\ref{fn:PMcorr}.} the IR causality condition \eqref{eq:grav_IRcaus_2} imposes the constraint $C_{\z_1} \simeq C_{\z_3}$ through spin refinement, although the condition itself cannot determine their allowed values.

\section{Discussions} \label{sec:conclusions}
Using scattering amplitudes, we explored how wave scattering on black hole backgrounds is affected by spin effects in EFTs of gravity based on the recently developed Magnusian formalism~\cite{Kim:2025gis}. As concrete targets, we studied the gravitational Faraday rotation, gravitational spin Hall effect, and time delay in EFTs of gravity parametrised by the action \eqref{The EFT}. We promoted the Magnusian to the Magnusian matrix (see sec.~\ref{sec:Magnusian}), and also defined the time delay as an observable derived from the Magnusian as \eqref{def: Time delay}, which can be viewed as the classical limit of the Smith time delay~\cite{Smith:1960zza}. 

We observed that the EFT corrections qualitatively alter the gravitational spin Hall effect of photons and gravitons; instead of splitting a ray of light into two distinct rays as in the Stern--Gerlach experiment, the ray of light is dispersed into a blob because the wavenumber kick (or the deflection angle) along orthogonal directions cannot be simultaneously diagonalised; see sec.~\ref{sec:observables1}. Another observation is that the spin effects can be tuned to refine the causality bounds based on ``positivity'' of the time delay, often leading to small $\CO(0.1)$ improvements on the bounds compared to the non-spinning case. A particularly intriguing refinement on causality bounds comes from considering graviton scattering with $R^4$ corrections, where initially uncorrelated Wilson coefficients were shown to become correlated when we start to consider spin effects; see fig.~\ref{fig:R4caus}. It would be interesting to understand if this correlation is an artefact of truncating the Magnusian at low orders in perturbation theory, or an effect that persists non-perturbatively.

In closing, we comment on some future directions regarding derivation of positivity bounds from multiparticle kinematics. 
Although our causality constraint studies in sec.~\ref{sec:TD_causality} are based on the time delay computed from $2 \to 2$ massive--massless scattering amplitudes, our definition of the time delay \eqref{def: Time delay} also applies to multiparticle kinematics (e.g. $3 \to 3$ scattering) and opens up a possibility of deriving causality constraints from multi-leg amplitudes that cannot be obtained from $2 \to 2$ subprocesses. 
Another extension along this direction would be to consider ``multipositivity bounds'' on the $\chi$ matrix. 
The multipositivity bounds derived in ref.~\cite{Cheung:2025nhw} rely on interpreting the residues of tree-level scattering amplitudes as expectation values of operators, and the bounds follow from the positivity of the operators' moments such as their variance ($\langle \CO_n^2 \rangle - \langle \CO_n \rangle^2 \ge 0$). 
This mechanism applies more naturally to the $\chi$-matrix given as the logarithm of the $S$-matrix ($S = e^{i \chi}$): 
the quantum version of the scattering generator equation \eqref{eq:scgendef}~\cite{Kosower:2018adc,Damgaard:2021ipf,Damgaard:2023ttc}
\begin{align}
\begin{aligned}
    \langle O \rangle_{\text{out}} = \langle \psi_{\text{out}} | O | \psi_{\text{out}} \rangle &= \langle \psi_{\text{in}} | e^{-\frac{i}{\hbar} \chi} \, O \, e^{\frac{i}{\hbar} \chi} | \psi_{\text{in}} \rangle = \langle \psi_{\text{in}} | e^{\frac{1}{i \hbar} [\chi , \bullet]} [O] | \psi_{\text{in}} \rangle
    \\ &= \langle \psi_{\text{in}} | \,  O + \frac{1}{i\hbar} [\chi , O] + \frac{1}{2!} \frac{1}{(i\hbar)^2} [ \chi , [ \chi , O]] + \cdots \, | \psi_{\text{in}} \rangle \,,
\end{aligned} \label{eq:qscgen}
\end{align}
computes the expectation value of an operator after the scattering process. Reorganising \eqref{eq:qscgen} to compute postive moments will lead to positivity conditions on the matrix elements of $\chi$, which could be translated to positivity conditions on the Wilson coefficients of the theory. 
Considering that studies on positivity bounds from multiparticle kinematics are rather scarce~\cite{Chandrasekaran:2018qmx,Arkani-Hamed:2023jwn,Guerrieri:2024ckc,Bresciani:2025toe,Cheung:2025nhw,Berman:2025owb,Elvang:2026pmc,Saha:2026ftv,Basile:2026gnd,Cheung:2026lpv,Jeong:2026xzk}, it would be interesting to understand how the $\chi$ matrix allows us to systematically explore the constraints arising from multiparticle kinematics.

\acknowledgments
JWK appreciates insightful discussions with Dean Carmi, Mariana Carrillo Gonz\'{a}lez, Stefano de Angelis, Yu-Tin Huang, Shota Komatsu, and Alexander Zhiboedov. KJ and SL thank Dongmin Gang for helpful discussions. KJ and SL are supported in part by the National Research Foundation of Korea (NRF) grant NRF-2022R1C1C1011979. KJ and SL also acknowledge support from the National Research Foundation of Korea (NRF) grant RS-2024-00405629. 
\appendix

\section{Evaluating spinor brackets} \label{app:spinors}
We use spinor conventions of ref.~\cite{Chung:2018kqs}. Spinor components are given as
\begin{align}
    |k\rg_{\a}\overset{\cdot}{=}\frac{1}{\sqrt{k_t+k_z}}\begin{pmatrix}
        -k_x+ik_y\\
        k_t+k_z
    \end{pmatrix}\;,\quad [k|_{\da}\overset{\cdot}{=}\frac{1}{\sqrt{k_t+k_z}}\begin{pmatrix}
        -k_x-ik_y\\
        k_t+k_z
    \end{pmatrix}\;.
\end{align}
Especially, we can check for $k_3=k_2+q$ and $\vec{k}_2=\w\hat{z}$ in the Compton amplitude's frame,
\begin{align}
    \begin{split}
        \left.\lg k_2k_3\rg\right|_{k_3=k_2+q}&=-\frac{\bfq}{\sqrt{1+\frac{q_z}{2\w}}}\sim -\bfq\;,
        \\ \left.[k_2k_3]\right|_{k_3=k_2+q}&=\frac{\fq}{\sqrt{1+\frac{q_z}{2\w}}}\sim \fq
    \end{split}
\end{align}
where holomorphic transfer momenta $\fq,\bfq$ are defined as $\fq\equiv q_x+iq_y$, $\bfq\equiv q_x-iq_y$. These factors enter the helicity flipping amplitudes.

\section{Seed amplitudes} \label{app:seed_amps}
We consider the action \eqref{The EFT}. All amplitudes are written in the all-incoming convention. For parity-even couplings, the parity-conjugate/helicity-reversed amplitudes can be obtained by exchanging angle and square spinors.

\subsection{3-point amplitudes}
Two-photon--one-graviton amplitudes:
\begin{align}
    M_{\text{EHM}}(1^{+},2^{-2},3^{+2})&=-\frac{[13]^4}{[12]^2}\;,
    \\M_{FFR}(1^{+},2^{+},3^{+2})&=\frac{\b}{4}\times[13]^2[23]^2\;.
\end{align}
Three graviton amplitudes:
\begin{align}
    M_{\text{EHM}}(1^{+2},2^{+2},3^{-2})&=-\frac{[12]^6}{[13]^2[23]^2}\;,
    \\ M_{R^3}(1^{+2},2^{+2},3^{+2})&=-\frac{\a'^2}{16}\times[12]^2[23]^2[31]^2\;.
\end{align}

\subsection{4-point amplitudes}
Mandelstam variables:
\begin{align}
    s = (k_1 + k_2)^2 \,,\quad t = (k_1 + k_4)^2 \,,\quad u = (k_1 + k_3)^2 \;.
\end{align}
Two-photon--two-graviton amplitudes:
\begin{align}
    M_{\text{EHM}}(1^{+},2^{+2},3^{-2},4^{-})&=\frac{[12]^4\lg 34\rg^2\lg13\rg^2 }{stu}\;,
    \\ M_{R^3}(1^+,2^{-2},3^{-2},4^-)&=-\frac{\a'^2}{16}\times\frac{\lg23\rg^4[1|2|4\rg^2}{t}\;,
    \\M_{FFR}(1^-,2^{+2},3^{-2},4^-)&=-\frac{\b}{4}\times\frac{\lg14\rg^2[2|1|3\rg^4}{stu}\;,
    \\M_{FFR}(1^{-},2^{-2},3^{-2},4^-)&=\frac{\b}{4}\times\bigr(\frac{\lg 12\rg^2\lg23\rg^2\lg34\rg^2}{s}+\frac{\lg 42\rg^2\lg23\rg^2\lg31\rg^2}{u}\bigr)\;.
\end{align}
Four-graviton amplitudes: 
\begin{align}
        M_{\text{EH}}(1^{+2},2^{+2},3^{-2},4^{-2})&=\frac{[12]^4\lg34\rg^4}{stu}\;,
        \\M_{R^3}(1^{+2},2^{-2},3^{-2},4^{-2})&=-\frac{\a'^2}{16}\times\bigr([12]\lg24\rg[41]\bigr)^2\frac{\lg23\rg\lg34\rg\lg42\rg}{[23][34][42]}\;,
        \\M_{R^3}(1^{+2},2^{+2},3^{+2},4^{+2})&=\frac{\a'^2}{4}\times stu\frac{[12][23][34][41]}{\lg12\rg\lg23\rg\lg34\rg\lg41\rg}\;,
        \\M_{R^4}(1^{+2},2^{+2},3^{-2},4^{-2})&=\tilde{\z}\times [12]^4\lg34\rg^4\;,
        \\M_{R^4}(1^{+2},2^{+2},3^{+2},4^{+2})&=\z^-\times\bigr([12]^4[34]^4+[13]^4[24]^4+[14]^4[23]^4\bigr)\;,
    \\M_{R^4}(1^{-2},2^{-2},3^{-2},4^{-2})&=\zeta^+\times\bigr(\lg12\rg^4\lg34\rg^4+\lg13\rg^4\lg24\rg^4+\lg14\rg^4\lg23\rg^4\bigr) \;,
\end{align}
where $\tilde{\z}\equiv 4(\z_1+\z_3)$ and $\zeta^\pm\equiv 4(\z_1-\z_3\pm\frac{i}{2}\z_2)$. The $\z^\pm$ couplings are parity-odd, therefore the last two amplitudes should be considered separately.

\subsection{Black hole factors}
The black hole factors given as shaded blobs in the main text are reproduced from ref.~\cite{Chen:2021kxt}. The expressions are expanded to linear order in spin in computations.

BH-BH-graviton amplitude, graviton helicity $h = \pm 2$:
\begin{align}
    C_3^{h} &= \frac{\k m x^h }{2} \exp \left( -i \frac{k_3^\m \e_3^\n S_{\m\n}}{p_1\cdot \e_3} \right)\,,\quad x = \frac{[3|p_1|\xi\rg}{m\lg 3\xi\rg} \,,\quad \e_3^\m = \frac{[3|\bs^\m|\xi\rg}{\sqrt{2}\lg 3\xi\rg}\;.
\end{align}
$|\xi\rg$ is an auxiliary spinor and $x$ is the proportionality factor introduced in ref.~\cite{Arkani-Hamed:2017jhn}. The spin tensor $S^{\m\n}$ is related to the spin vector $S^\m$ by the relation $S_{\m\n} = -\frac{1}{m}\e_{\m\n\r\s}p_1^\r S^\s$.

BH--BH--two-graviton amplitude, graviton helicity \(h,h'=\pm 2\):
\begin{align}
    C_4^{hh'}=\frac{\mathcal{S}^{hh'}}{t(s-m^2)(u-m)^2}\times\begin{cases}
    [2|p_1|3\rg^4\,, & (h,h')=(+2,-2)
    \\m^4[23]^4\,,&(h,h')=(+2,+2)
    \end{cases}
\end{align}
\(\mathcal{S}^{hh'}\) is the spin factor \(\mathcal{S}^{hh'}=\exp(iK^\mu A^{hh'}_\mu)\) where \(K^\mu = l_2^\mu + l_3^\mu\) and~\cite{Chen:2021kxt,Chung:2018kqs}
\begin{align}
    A_\mu^{+-}= L^\nu S_{\mu\nu}\,, \quad A_\mu^{++} =  \frac{if(t/m^2)}{m}S_\mu\,,\quad L^\m= -\frac{[2|\bs^\m|3\rg}{[2|p_1|3\rg}\,,\quad f(x) = \sum_{n=0}^\infty \frac{(n!)^2 x^n}{(2n+1)!} \,.
\end{align}

\bibliographystyle{JHEP}
\bibliography{main,ref_noinspire}

\end{document}